\pdfoutput=1
\documentclass[atlasdraft=false, texlive=2023, UKenglish, block=none, cernpreprint, texmf, orcidlogo]{atlasdoc}
 
\usepackage{atlaspackage}
\usepackage{atlasbiblatex}
\usepackage{afterpage,refcount,lastpage}
\usepackage{lscape}
\usepackage{booktabs}
\usepackage{siunitx}
\usepackage{tablefootnote}
\usepackage{atlasphysics}

\graphicspath{{logos/}{figures/}}
 
\usepackage{ANA-HDBS-2019-16-PAPER-defs}

\AtlasTitle{A search for heavy Higgs bosons decaying into vector bosons in same-sign two-lepton final states in $pp$ collisions at
$\sqrt{s}=13$~TeV with the ATLAS detector}
 
\AtlasAbstract{
A search for heavy Higgs bosons produced in association with a vector boson and decaying into a pair of vector bosons is performed in final states with two leptons (electrons or muons) of the same electric charge, missing transverse momentum and jets. A data sample of proton--proton collisions at a centre-of-mass energy of 13~TeV recorded with the ATLAS detector at the Large Hadron Collider between 2015 and 2018 is used. The data correspond to a total integrated luminosity of $139~\mathrm{fb^{-1}}$. The observed data are in agreement with Standard Model background expectations. The results are interpreted using higher-dimensional operators in an effective field theory. Upper limits on the production cross-section are calculated at 95\% confidence level as a function of the heavy Higgs boson's mass and coupling strengths to vector bosons. Limits are set in the Higgs boson mass range from 300 to 1500~GeV, and depend on the assumed couplings. The highest excluded mass for a heavy Higgs boson with the coupling combinations explored is 900~GeV. Limits on coupling strengths are also provided.
 
}

\author{The ATLAS Collaboration}
 
\AtlasRefCode{HDBS-2019-16}

\AtlasJournal{JHEP}
\AtlasJournalRef{JHEP 07 (2023) 200}
\AtlasDOI{10.1007/JHEP07(2023)200}
\PreprintIdNumber{CERN-EP-2022-187}

\hypersetup{pdftitle={ATLAS document},pdfauthor={The ATLAS Collaboration}}
 
\newif\ifconf
\conffalse
\begin{document}
 
\maketitle
 
\tableofcontents
\newpage


\section{Introduction}
 
The discovery of the Higgs boson was a major success for the Standard Model (SM) and an important breakthrough in understanding
electroweak symmetry breaking~\cite{HIGG-2012-27,CMS-HIG-12-028}. It also opened new ways to search for physics beyond the Standard Model (BSM physics).
Despite its success the SM is not without problems that may require extensions and new concepts.
One natural extension common to many BSM physics models is an extended Higgs sector, which leads to the introduction of additional Higgs bosons.
In such models, the lightest Higgs boson ($h$) often has properties similar to those of the observed SM Higgs boson.
Additional Higgs bosons have been introduced to explain a very wide range of BSM phenomena, from the observed baryon asymmetry in the universe, and how to solve the strong CP problem with the help of axions, to the generation of non-zero neutrino masses~\cite{Branco:2011iw, Bhattacharyya:2015nca, Aggleton:2016tdd, Davoudiasl:2022mav, Wang:2022yhm}.
Some models have incorporated recent potential deviations from the SM seen in muon $g-2$ and $W$ mass measurements~\cite{Kim:2022hvh}.

Several searches for additional heavy Higgs bosons ($H$) have already been carried out with the ATLAS and CMS detectors at the Large Hadron Collider (LHC)~\cite{
HDBS-2019-06,HIGG-2018-09,HDBS-2018-10,HIGG-2016-09,HIGG-2016-31, CMS-HIG-17-033,CMS-HIG-18-005}. These searches mainly relied on the gluon--gluon fusion (ggF) and vector-boson fusion (VBF) production mechanisms, which are the dominant production modes for the SM Higgs boson at the LHC. In ggF production, the gluons couple to the Higgs boson mainly via a top-quark loop, so the non-observation of an additional Higgs boson in this channel could point to a reduced fermionic coupling.

This analysis concentrates on the production of a heavy Higgs boson in association with a vector boson ($VH$, where $V=W,Z$) and $H\to VV$ decays.
By utilising a general effective Lagrangian that includes dimension-six operators in an effective field theory (EFT), it can be shown that, relative to VBF production, the $VH$ production mode benefits from having smaller SM backgrounds, and the production cross-section may be enhanced by higher-order contributions, especially at high Higgs boson mass and vector-boson momenta~\cite{Ren:2014bza,Kuang:2015gka,Chen:2019hey}.
The observed Higgs boson $h$, is assumed to have the production and decay modes as in the SM, with production through $H\to Zh$ negligible.
 
\begin{figure}[!htb]
\begin{center}
\includegraphics[width=0.85\textwidth]{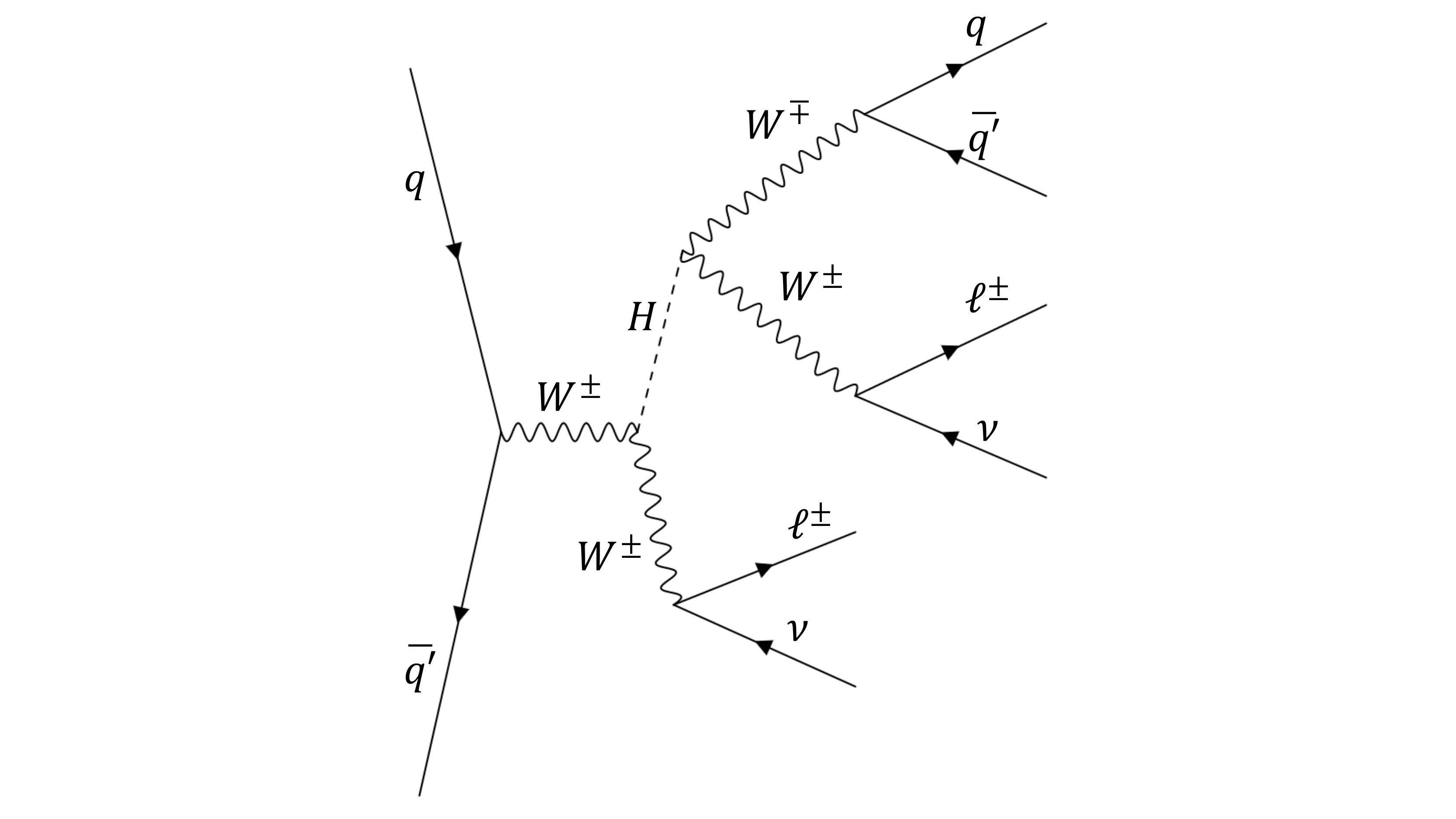}
\caption{Feynman diagram of the $W^{\pm}H \rightarrow \www$ process.}
\label{fig:Feyman}
\end{center}
\end{figure}

Rather than focusing on any specific model, a generic search is performed for the same-sign dilepton signature (SS2L), targeting the $W^{\pm}H \rightarrow \www\rightarrow \ell^\pm\nu\ell^\pm\nu jj$ decay channel. The corresponding Feynman diagram is shown in Figure~\ref{fig:Feyman}. In this article, the term `lepton', unless stated otherwise, refers to either an electron or a muon. Electrons and muons from $\tau$-lepton decays are also considered.
The hadronically decaying $W$ boson is reconstructed either as two small-radius jets or as a single large-radius jet for higher-momentum $W$ bosons.
The presence of neutrinos prevents the full reconstruction of the heavy Higgs boson's mass and is the main drawback of the $\www$ channel with leptonic decays of the $W$ bosons.
It is ameliorated with the help of the reconstruction methods described in Section~\ref{sec:selections}.
Compared with other bosonic $VH$ decay channels, the chosen channel has the highest signal sensitivity thanks to low SM backgrounds and a sizeable branching fraction for
$H\to WW$ decay~\cite{Chen:2019hey, Heinemeyer:2013tqa}.
SM processes produce same-sign lepton pairs at the LHC at a rate that is orders of magnitude below that of opposite-sign lepton-pair production in the SM.

\section{Phenomenology}
\label{sec:phenomenology}
\newcommand{\ctw}{c_W}
\newcommand{\stw}{s_W}
In theories with multiple Higgs fields, the fields in the multi-Higgs potential interact and the mass eigenstates are formed from a mixture of the fields.
In the simple case of a lightest ($h$) and next-to-lightest ($H$) neutral Higgs doublet, the couplings of the Higgs boson to the SM gauge bosons are scaled relative to SM gauge couplings because of the mixing.
In addition to the leading-order dimension-four (dim-4) operators, dimension-six (dim-6) effective operators as described in Refs.~\cite{Ren:2014bza,Kuang:2015gka,Chen:2019hey} are also considered. After electroweak symmetry breaking, the effective Lagrangian terms can be written as
\begin{align*}
\mathcal{L}^{(4)}_{hWW} & = \rho_{h} gm_{W} hW^{\mu}W_{\mu}, \\
\mathcal{L}^{(4)}_{hZZ} & = \rho_{h} \frac{gm_{W}}{2\ctw^{2}} hZ^{\mu}Z_{\mu}, \\
\mathcal{L}^{(4)}_{HWW} & = \rho_{H} gm_{W} HW^{\mu}W_{\mu}, \\
\mathcal{L}^{(4)}_{HZZ} & = \rho_{H} \frac{gm_{W}}{2\ctw^{2}} HZ^{\mu}Z_{\mu},\\
\mathcal{L}^{(6)}_{HWW} & = \rho_{H} gm_{W}\frac{f_{W}}{2\Lambda^{2}} \left( W^{+}_{\mu\nu}W^{-\mu}\partial^{\nu}H +h.c. \right) - \rho_{H} gm_{W}\frac{f_{WW}}{\Lambda^{2}} W^{+}_{\mu\nu}W^{-\mu\nu}H, \\
\mathcal{L}^{(6)}_{HZZ} & = \rho_{H} gm_{W}\frac{\ctw^{2}f_{W}+\stw^{2}f_{B}}{2\ctw^{2}\Lambda^{2}} Z_{\mu\nu}Z^{\mu}\partial^{\nu}H - \rho_{H} gm_{W}\frac{\ctw^{4}f_{WW}+\stw^{4}f_{BB}}{2\ctw^{2}\Lambda^{2}} Z_{\mu\nu}Z^{\mu\nu}H,
\end{align*}
 
where  $h,H, W$ and $Z$ are the fields of the light and heavy Higgs bosons and the $W$ and $Z$ bosons, respectively; $m_W$ is the $W$ boson mass; $g$ is the SM coupling constant of the weak interaction; $f_{W}$,  $f_{WW}$, $f_{B}$,  and $f_{BB}$ are anomalous couplings to $W$ and $B$ fields; $\rho_h$ and $\rho_H$ are scaling factors; $\stw = \sin \theta_{W}$ and $\ctw = \cos \theta_W$, where $\theta_W$ is the weak mixing angle; and $\Lambda$ is the scale below which the effective Lagrangian holds. For a light Higgs boson similar to the one in the SM, $\rho_{h}$ is close to 1. The simplest two-Higgs-doublet model (2HDM)~\cite{Gunion:1989we} has $\rho_{h} = \cos(\beta-\alpha)$ and $\rho_{H} = \sin(\beta-\alpha)$, where $\alpha$ is the mixing angle between the CP-even Higgs bosons, and $\beta$ is the rotation angle, with $\tan \beta$ defined as the ratio of vacuum expectation values of the two Higgs doublets. In this analysis, the scaling factor $\rho_{H}$ is set to $0.05$ and the scale $\Lambda$ is set to $5~\TeV$, which is much larger than the mass of the heavy Higgs boson in this search. The choice of $\rho_{H}$ is motivated by the observation that $\rho_{h}\sim 1$. To further simplify the parameter space, the small terms of $O(\stw^{2})$ and $O(\stw^{4})$ are neglected, and the anomalous coupling coefficients $f_{B}$ and $f_{BB}$ are set to zero, following Ref. \cite{Ren:2014bza}. The operator multiplied by the $f_{W}$ anomalous coupling is proportional to derivatives of the Higgs field and hence production is enhanced with increasing Higgs boson momentum.
Results for the heavy Higgs boson's production cross-section are provided as a function of the heavy Higgs boson's mass and the two BSM $HVV$ coupling strengths $\rho_{H}f_{W}/\Lambda^{2}$ and $\rho_{H}f_{WW}/\Lambda^{2}$.


\section{ATLAS detector and data samples}
 
The ATLAS detector~\cite{PERF-2007-01} at the LHC covers nearly the entire solid angle around the collision point.\footnote{ATLAS uses a right-handed coordinate system with its origin at the nominal interaction point (IP) in the centre of the detector and the $z$-axis coinciding with the axis of the beam pipe.  The $x$-axis points from the IP towards the centre of the LHC ring, and the $y$-axis points upward. Cylindrical coordinates ($r$,$\phi$) are used in the transverse plane, $\phi$ being the azimuthal angle around the $z$-axis. The pseudorapidity is defined in terms of the polar angle $\theta$ as $\eta = - \ln \tan(\theta/2)$. The distance in ($\eta$,$\phi$) coordinates, $\Delta R = \sqrt{(\Delta\phi)^2+(\Delta\eta)^2}$, is also used to define cone sizes. Rapidity is defined as $y=(1/2) \ln[(E+p_z)/(E-p_z)]$, where $E$ is the energy and $p_z$ is the $z$-component of the momentum. Transverse momentum and energy are defined as $\pt=p\sin\theta$ and $\et=E\sin\theta$, respectively.}
It consists of an inner tracking detector surrounded by a thin superconducting solenoid, electromagnetic and hadron calorimeters,
and a muon spectrometer incorporating three large superconducting air-core toroidal magnets.
 
The inner-detector system (ID) is immersed in a \SI{2}{\tesla} axial magnetic field
and provides charged-particle tracking in the range \(|\eta| < 2.5\).
The high-granularity silicon pixel detector covers the vertex region and typically provides four measurements per track,
the first hit normally being in the insertable B-layer (IBL) installed before Run~2~\cite{ATLAS-TDR-19, PIX-2018-001}.
It is followed by the silicon microstrip tracker (SCT), which usually provides eight measurements per track.
These silicon detectors are complemented by the transition radiation tracker (TRT),
which enables radially extended track reconstruction up to \(|\eta| = 2.0\).
The TRT also provides electron identification information
based on the fraction of hits (typically 30 in total) above a higher energy-deposit threshold corresponding to transition radiation.
 
The calorimeter system covers the pseudorapidity range \(|\eta| < 4.9\).
Within the region \(|\eta|< 3.2\), electromagnetic calorimetry is provided by barrel and
endcap high-granularity lead/liquid-argon (LAr) calorimeters,
with an additional thin LAr presampler covering \(|\eta| < 1.8\)
to correct for energy loss in material upstream of the calorimeters.
Hadron calorimetry is provided by the steel/scintillator-tile calorimeter,
segmented into three barrel structures within \(|\eta| < 1.7\), and two copper/LAr hadron endcap calorimeters.
The solid angle coverage is completed with forward copper/LAr and tungsten/LAr calorimeter modules
optimised for electromagnetic and hadronic energy measurements, respectively.
 
The muon spectrometer (MS) comprises separate trigger and
high-precision tracking chambers measuring the deflection of muons in a magnetic field generated by the superconducting air-core toroidal magnets.
The field integral of the toroids ranges between \num{2.0} and \SI{6.0}{\tesla\metre}
across most of the detector.
Three layers of precision chambers, each consisting of layers of monitored drift tubes, covers the region \(|\eta| < 2.7\),
complemented by cathode-strip chambers in the forward region, where the background is highest.
The muon trigger system covers the range \(|\eta| < 2.4\) with resistive-plate chambers in the barrel, and thin-gap chambers in the endcap regions.
 
Interesting events are selected by the first-level trigger system implemented in custom hardware,
followed by selections made by algorithms implemented in software in the high-level trigger~\cite{TRIG-2016-01}.
The first-level trigger accepts events from the \SI{40}{\MHz} bunch-crossings at a rate below \SI{100}{\kHz},
which the high-level trigger further reduces in order to record events to disk at about \SI{1}{\kHz}.

The data used in this analysis were collected using single-lepton triggers
during the 2015--2018 proton--proton ($pp$) collision running periods at a centre-of-mass energy of 13~\TeV.
Events are selected for analysis only if they are of good quality and if all the relevant
detector components are known to have been in good operating condition, which corresponds to a total integrated luminosity of
$139$~\fbi~\cite{ATLAS-CONF-2019-021,DAPR-2018-01}.
The recorded events contain an average of 34 inelastic $pp$ collisions per bunch-crossing.
 
An extensive software suite~\cite{ATL-SOFT-PUB-2021-001} is used in data simulation, in the reconstruction and analysis of real
and simulated data, in detector operations, and in the trigger and data acquisition systems of the experiment.


\section{Simulation of signal and background processes}
 
Monte Carlo (MC) simulated event samples are used to model heavy Higgs boson signals and to estimate the SM background with two same-sign leptons and/or at least three prompt leptons. Data-driven methods are used to estimate charge-flip, non-prompt and photon-conversion backgrounds, as discussed in Section~\ref{sec:backgrounds}.
 
The heavy Higgs boson signal process $pp\!\to\! VH\! \to\! VVV$ was modelled at leading order (\LO) in QCD by the \MGNLO[2.7.3] generator~\cite{Alwall:2014hca}.
The full decays of $V$ bosons were simulated in \MADSPIN~\cite{Artoisenet:2012st}.
Parton showers and hadronisation were handled by \PYTHIA[8.244]~\cite{Sjostrand:2014zea} using the A14 set of tuned parameters~\cite{ATL-PHYS-PUB-2014-021} and the \NNPDF[2.3lo]~\cite{Ball:2012cx} parton distribution function (PDF).
Events were filtered such that at least one same-sign lepton pair was produced.
Each lepton was required to have transverse momentum larger than $18~\GeV$, and be within $|\eta| < 2.7$.
The samples were produced with $m_H$ from $300~\GeV$ to $1.5~\TeV$, $f_W$ from $-2480$ to $2510$ and $f_{WW}$ from $-15000$ to $15000$.
The event samples are normalised to calculations at next-to-leading order (NLO) using a Higgs characterisation model~\cite{NLOHC}. The NLO $K$-factor increases the expected event yields by a factor of 1.3, independently of the heavy Higgs boson's mass and BSM $HVV$ coupling strengths.
 
Samples of diboson final states ($VV$) were simulated with the
\SHERPA[2.2.2]~\cite{Bothmann:2019yzt} generator,
including off-shell effects and Higgs boson contributions, where appropriate.
Fully leptonic final states were generated using
matrix elements at NLO accuracy in QCD for up to one additional parton
and at LO accuracy for up to three additional parton
emissions. Samples for the loop-induced processes $gg \to VV$ were
generated using LO-accurate matrix elements for up to one
additional parton emission for both the fully leptonic and
semileptonic final states. The matrix element calculations were matched
and merged with the \SHERPA parton shower based on Catani--Seymour
dipole factorisation~\cite{Gleisberg:2008fv,Schumann:2007mg} using the \MEPSatNLO
prescription~\cite{Hoeche:2011fd,Hoeche:2012yf,Catani:2001cc,Hoeche:2009rj}.
The virtual QCD corrections are provided by the
\OPENLOOPS library~\cite{Buccioni:2019sur,Cascioli:2011va,Denner:2016kdg}. The
\NNPDF[3.0nnlo] set of PDFs was used~\cite{Ball:2014uwa}, along with the
dedicated set of tuned parton-shower parameters developed by the
\SHERPA authors.
 
The production of triboson ($VVV$) events was simulated with the
\SHERPA[2.2.2] generator using factorised gauge-boson decays.
Matrix elements, accurate to NLO for the inclusive process and to LO for up to
two additional parton emissions, were matched and merged with the \SHERPA parton
shower based on Catani--Seymour dipole factorisation
using the \MEPSatNLO prescription.
The virtual QCD corrections for matrix elements at NLO accuracy are
provided by the \OPENLOOPS library.
Samples were generated using the \NNPDF[3.0nnlo] PDF set, along with
the dedicated set of tuned parton-shower parameters developed by the \SHERPA authors.
Contributions with an off-shell $W$ boson through $Wh \to WWW^*$ were generated using \POWHEGBOX[v2]~\cite{Frixione:2007nw,Nason:2004rx,Frixione:2007vw,Alioli:2010xd} interfaced to \PYTHIA[8.235] to model the parton shower with the \NNPDF[2.3lo] PDF and the \AZNLO set of tuned parameters~\cite{STDM-2012-23}.
 
The production of \tth\ events was modelled using the \POWHEGBOX[v2] generator at NLO with the \NNPDF[3.0nlo] PDF set. The events were interfaced to \PYTHIA[8.230] with parameters set according to the A14 tune and using the \NNPDF[2.3lo] set of PDFs. The decays of bottom and charm hadrons were performed by \EVTGEN[1.6.0]~\cite{Lange:2001uf}.
The production of \ttV,\ \tWZ\ and \tZq\ events was modelled using the
\MGNLO[2.3.3]~\cite{Alwall:2014hca} generator at NLO with the
\NNPDF[3.0nlo] PDF.
The events were interfaced to \PYTHIA[8.210], which
used the A14 tune and the
\NNPDF[2.3lo] PDF set. The decays of bottom and charm
hadrons were simulated using the \EVTGEN[1.2.0] program.
 
Events from \ttbar, $V$+\,jets and $V\gamma$ processes contribute to the $\ell^\pm\ell^\pm$ signal region when a lepton charge is mismeasured, or leptons are produced from non-prompt decays or photon conversions. These backgrounds are estimated with data-driven methods, as detailed in Section~\ref{sec:backgrounds}, with MC simulation used for validation and to estimate systematic uncertainties. These MC simulations are briefly introduced in the following paragraphs.

The production of \ttbar events was modelled by the
\POWHEGBOX[v2]
generator at NLO with the \NNPDF[3.0nlo] PDF set
and the \hdamp parameter\footnote{The
\hdamp parameter is a resummation damping factor and one of the
parameters that controls the matching of \POWHEG matrix elements to
the parton shower and thus effectively regulates the
high-\pt radiation against which the \ttbar system recoils.} set
to 1.5 times the top-quark mass, \mtop~\cite{ATL-PHYS-PUB-2016-020}.  The events were interfaced
to \PYTHIA[8.230] to model the parton showers,
hadronisation, and underlying event, with parameters set according
to the A14 tune and using the \NNPDF[2.3lo]
set of PDFs. The decays of bottom and charm hadrons
were performed by \EVTGEN[1.6.0]~\cite{Lange:2001uf}.
 
The production of $V$+\,jets events was simulated with the \SHERPA[2.2.1]
generator using NLO matrix elements for up to two partons, and LO matrix elements
for up to four partons calculated with the Comix and \OPENLOOPS libraries. They were matched with the \SHERPA parton shower using the \MEPSatNLO prescription and the set of tuned parameters developed by the \SHERPA authors.
The \NNPDF[3.0nnlo] set of PDFs was used and the samples
are normalised to a next-to-next-to-leading-order (NNLO)
prediction~\cite{Anastasiou:2003ds}.
 
The production of $V\gamma$ final states was simulated with the
\SHERPA[2.2.2] generator.  Matrix elements at
NLO accuracy in QCD for up to one additional parton and LO accuracy for
up to three additional parton emissions were matched and merged with
the \SHERPA parton shower as described for the diboson processes.

The generated events were passed through a simulation of the ATLAS detector and its response~\cite{SOFT-2010-01} based on \GEANT~\cite{Agostinelli:2002hh}, and reconstructed using the same software framework as for data~\cite{ATL-SOFT-PUB-2021-001}.
The effect of multiple interactions in the same and neighbouring bunch-crossings (\pileup)
was modelled by overlaying each simulated hard-scattering event with
inelastic $pp$ events generated with \PYTHIA[8.186]~\cite{Sjostrand:2007gs}
using the \NNPDF[2.3lo] set of PDFs and the A3 set of tuned parameters~\cite{ATL-PHYS-PUB-2016-017}.


\section{Object reconstruction and identification}
\label{sec:objects}
 
Proton--proton interaction vertices are reconstructed from charged-particle tracks with transverse momenta $\pT>500$~\MeV~\cite{ATL-PHYS-PUB-2015-026,PERF-2015-01}. The vertex with the highest sum of squared transverse momenta of associated tracks is selected as the primary vertex of the hard interaction.
 
Electrons are reconstructed from topological clusters of energy deposits in the electromagnetic calorimeter which are matched to a track in the inner detector~\cite{EGAM-2018-01}.
Electrons are required to have $\pT>20$~\GeV\ and to be reconstructed within $|\eta|<2.47$, excluding electrons in the transition region between the barrel and endcap calorimeters ($1.37<|\eta|<1.52$).
The electron identification is based on a multivariate likelihood-based discriminant that uses the shower shapes in the electromagnetic calorimeter and the associated ID track properties.
Electrons are required to satisfy the \textit{Tight} identification criterion for better rejection of non-prompt electrons~\cite{EGAM-2018-01}.
Muon candidates are identified by matching ID tracks to full tracks or track segments reconstructed in the muon spectrometer or by using only information from the muon spectrometer outside of the ID acceptance~\cite{MUON-2018-03}.
Muons are required to have $\pT>20$~\GeV, to be reconstructed within $|\eta|<2.5$, and to satisfy the \textit{Medium} cut-based identification criterion as defined in Ref.~\cite{PERF-2015-10}.
To have an origin compatible with the primary vertex, electrons (muons) must satisfy $|d_0/\sigma_{d_0}|<5~(3)$ and $|\zzsth|<0.5$~mm. Here $d_0$ is the transverse impact parameter,
$\sigma_{d_0}$ is its uncertainty, $z_0$ is the distance along the $z$-axis from the primary vertex to the point where $d_0$ is measured, and $\theta$ is the polar angle of the track.
In order to reject leptons likely to have originated from non-prompt hadronic decays, leptons are required to satisfy a
criterion based on ID and calorimeter isolation variables and the output of a boosted-decision tree (BDT) in a prompt-lepton-veto tagger algorithm~\cite{HIGG-2017-02}. Electrons must also pass a \textit{charge misidentification suppression BDT} which rejects electrons likely to have a wrongly measured electric charge~\cite{EGAM-2018-01}.
Furthermore, in order to reduce the number of electrons likely to have originated from photon conversion, additional requirements are applied to the electron candidate (referred to as `photon-conversion electron suppression requirements')~\cite{EGAM-2018-01, TOPQ-2018-05}. It must not be associated with a reconstructed photon-conversion vertex in the detector material nor have a reconstructed displaced vertex with radius $r>20$ mm whose reconstruction uses the track associated with the electron. Finally the electron candidate's track and the closest track in $\Delta R$ at the primary vertex or a conversion vertex must not have an invariant mass below $100$~\MeV.

The anti-$k_t$ algorithm~\cite{Cacciari:2008gp,Cacciari:2011ma} with a radius parameter of 0.4 is used to reconstruct small-radius (small-$R$) jets up to $|\eta|=4.9$. It uses particle-flow objects, which combine tracking and calorimetric information, as input~\cite{PERF-2015-09}. Energy- and $\eta$-dependent correction factors derived from MC simulations are applied in order to correct jets back to the particle level~\cite{PERF-2016-04}. Jets are required to have $\pT>20$~\GeV\ and $|\eta|<2.5$.
To suppress jets from pile-up, a jet vertex tagger~\cite{PERF-2014-03} applied to jets with $\pT<120$~\GeV\ and $|\eta|<2.5$ uses information about tracks associated with the primary vertex and pile-up vertices. Jets containing $b$-flavoured hadrons are identified in the region $|\eta|<2.5$ with a $b$-tagging algorithm based on a recurrent neural network~\cite{FTAG-2018-01}. It makes use of the impact parameters of tracks associated with the jet, the position of reconstructed secondary vertices and their compatibility with the decay chains of such hadrons. At the chosen working point, the $b$-tagging algorithm provides light-flavour ($u$,$d$,$s$-quark and gluon) and $c$-jet rejection factors of 33 and 3, respectively, for an average 85\% $b$-jet tagging efficiency, as estimated from simulated $\ttbar$ events~\cite{FTAG-2018-01}.
 
Hadronically decaying $\tau$-leptons are reconstructed~\cite{PERF-2014-06,ATLAS-CONF-2017-029} as jets by applying the anti-$k_t$ algorithm with a radius parameter of 0.4 to noise-suppressed energy clusters. They are required to have exactly one or three tracks in the ID within a cone of size $\Delta R=0.2$ around the jet axis, to have $\pT > 20$~\GeV\ and $|\eta|<2.5$, and to be outside the transition region between the barrel and endcap calorimeters ($1.37<|\eta|<1.52$). To prevent jets from being reconstructed and misidentified as $\tau$-leptons, a multivariate approach using boosted decision trees, based on information from the calorimeters and tracking detectors, is employed. The `medium' quality criteria described in Ref.~\cite{ATLAS-CONF-2017-029} are applied. Hadronically decaying $\tau$-leptons are only used in the analysis in the overlap-removal procedure described at the end of this section.
 
Large-radius (large-$R$) jets are reconstructed from noise-suppressed topological clusters (topoclusters) of calorimeter energy depositions~\cite{PERF-2014-07}, using the anti-$k_t$ algorithm with a radius parameter of 1.0, with the topoclusters calibrated at the local hadronic scale~\cite{PERF-2014-07}. Large-$R$ jets are groomed using trimming~\cite{Krohn:2009th,PERF-2012-02} to improve the jet mass resolution and its stability with respect to pile-up by discarding the softer components of jets that originate from initial-state radiation, pile-up interactions, or the underlying event. Large-$R$ jets are required to have $\pT>200$~\GeV\ and $|\eta|<2.0$. Only large-$R$ jets with a jet mass $\mJ$ between $50$~\GeV\ and $200$~\GeV\ are considered in the analysis.
 
The missing transverse momentum (\metvec) is defined as the negative vector sum of the transverse momenta of electrons, muons, hadronically decaying $\tau$-leptons and small-$R$ jets in the event, plus a `soft-term' built from additional tracks associated with the primary vertex~\cite{PERF-2016-07,ATLAS-CONF-2018-023}. The magnitude of \metvec\ is denoted by $\met$.
 
An overlap-removal procedure is applied to the selected leptons and jets. Any hadronically decaying $\tau$-lepton reconstructed closer than $\Delta R=0.2$ to an electron or muon is removed.
Electrons that fall within $\Delta R=0.2$ of a selected muon are also discarded. For electrons and nearby small-$R$ jets, the jet is removed if the separation between the electron and jet satisfies $\Delta R<$ 0.2; the electron is removed if the separation satisfies $0.2<\Delta R<0.4$. For muons and nearby small-$R$ jets, the jet is removed if the separation between the muon and jet satisfies $\Delta R<$ 0.2 and the jet has less than three tracks or the energy and momentum differences between the muon and the jet are small; otherwise the muon is removed if the separation satisfies $\Delta R<$ 0.4. Small-$R$ jets that are reconstructed within a cone of size $\Delta R=0.2$ around the axis of a hadronically decaying $\tau$-lepton are removed. To prevent double-counting of energy from an electron inside the large-$R$ jet, the large-$R$ jet is removed if the separation between the electron and the large-$R$ jet satisfies $\Delta R < 1.0$.


\section{Event classification and selection}
\label{sec:selections}
 
The experimental signature of the $\ell^\pm\nu\ell^\pm\nu qq$ signal process requires the presence of two same-sign leptons, $\met$, and depending on the reconstruction of the $qq$ final state, two small-$R$ jets or one large-$R$ jet with an invariant mass close to 80~\GeV. The selection requirements used to define the signal regions are 
optimised to maximise the sensitivity to the $\ell^\pm\nu\ell^\pm\nu qq$ signal process while reducing contributions from SM background processes.
 
Events are required to satisfy a logical OR of single-electron~\cite{TRIG-2018-05} and single-muon~\cite{TRIG-2018-01} triggers with $\pT$ thresholds ranging from $20$~\GeV\ to $26$~\GeV\ and increasing from 2015 to 2018. All events must contain at least one lepton with $\pT>27$~\GeV\ that triggered the event, which ensures that the trigger efficiency reached its plateau. Events are required to have exactly two nominal leptons meeting the object criteria described in Section~\ref{sec:objects} and they must have the same electric charge. The invariant mass of the dilepton system is required to be larger than 100~\GeV\ to reduce the contributions coming from the $Z$\,+\,jets process. To reduce the SM background contributions from processes that have more than two leptons, a `veto lepton' definition is introduced. Compared with the nominal lepton selection criteria, the veto electron (muon) $\pT$ threshold is lowered to $7~(4.5)$~\GeV, and the isolation, charge misidentification suppression BDT, and photon-conversion electron suppression requirements are removed. For veto electrons, the \textit{Loose} likelihood-based identification definition is used~\cite{EGAM-2018-01}. For veto muons, the \textit{Loose} cut-based identification definition is used~\cite{MUON-2018-03}. Events with additional veto leptons are removed.
 
For the hadronically decaying $W$ boson, the energy deposits of the two resulting jets are either well separated or can largely overlap in the detector, depending on the momentum of the parent boson. Thus the $\Wqq$ decay can either be reconstructed from two resolved small-$R$ jets ($\Wjj$) for low-momentum bosons or identified as one merged large-$R$ jet ($\WJ$) for higher momentum, boosted bosons. An event is assigned to the boosted category if it contains at least one large-$R$ jet satisfying the object criteria described in Section~\ref{sec:objects}; otherwise the event is assigned to the resolved category. In turn, two signal regions are defined: the boosted signal region (boosted SR) and the resolved signal region (resolved SR).
 
In the boosted SR, the large-$R$ jet with the highest $\pT$ is selected as the candidate for the hadronically decaying $W$ boson and must satisfy $\pT>200$~\GeV. A boson tagger is subsequently applied to distinguish between jets from hadronically decaying $W$ bosons (which decay to two partons), and jets originating from a single quark or gluon~\cite{ATL-PHYS-PUB-2020-017}.
In this analysis, the boson tagger is configured to have 80\% identification efficiency for the hadronically decaying $W$ boson. An $\met>80$~\GeV\ selection is applied in this region.
 
In the resolved SR, at least two small-$R$ jets with $\pT>20$~\GeV\ and $\abseta<2.5$ are required. The invariant mass of the dijet system, formed by the two small-$R$ jets with the largest $\pT$, is required to be consistent with the $W$ boson mass: $50~\GeV <\mjj< 110$~\GeV. In both the boosted and resolved SRs, the events are required to have no $b$-tagged small-$R$ jet present to reduce the background from top-quark production. An $\met>60$~\GeV\ selection is applied in this region.
 
One of the dominant sources of SM background in the SRs is the $WZ$\,+\,jets process. Control regions enriched in this process ($WZ$ CRs) are defined for both the boosted and resolved categories, and are used in the global likelihood fit as detailed in Section~\ref{sec:results}, in order to constrain the normalisation of this background. The $WZ$ CRs require events with three leptons, of which two form a same-flavour opposite-sign (SFOS) pair. Similarly to the SRs, the $WZ$ CRs veto any events with $b$-jets to reduce backgrounds coming from top-quark processes and they veto events containing a fourth veto lepton to reduce background coming from the $ZZ$ process. Requirements of $\met>40$~\GeV\ and trilepton invariant mass $\mlll>100$~\GeV\ are applied in order to further reduce background coming from Drell--Yan and $ZZ$ processes. In the $WZ$ CRs, more than 90\% of the events are expected to be from $WZ$\,+\,jets production.
 
The same-sign $WW$\,+\,jets process is another important SM background in the SRs. A control region enriched with this background is defined (ss$WW$ CR) in the resolved category by requiring a dijet invariant mass $\mjj>200$~\GeV, and is also used in the global likelihood fit as detailed in Section~\ref{sec:results} to constrain the normalisation of this background. The $\mjj$ requirement enhances the fraction of same-sign $WW$\,+\,jets events from electroweak VBF production, as the two forward jets tend to have a large invariant mass~\cite{STDM-2017-06}. In order to reduce the statistical uncertainties, the $\met$ requirement is loosened to $\met>40$~\GeV. In the ss$WW$ CR, about 40\% of the events are expected to originate from same-sign $WW$\,+\,jets production.
 
The event selection criteria for the signal and control regions are summarised in Table~\ref{tab:selections}.
The average product of acceptance times efficiency for signal events in the combined SR is roughly 0.2\%--0.5\%, with little variation over the probed mass range for a heavy Higgs boson.

\renewcommand\arraystretch{1.3}
\begin{table}[ptbh]
\centering
\caption{Overview of the event selection criteria for the signal and control regions.}
\resizebox{\textwidth}{!}{
\begin{tabular}{l|c|c|c|c|c}
\toprule
\textbf{Selections} & Boosted SR & Resolved SR & $ssWW$ CR & Boosted $WZ$ CR & Resolved $WZ$ CR\\
\hline
Trigger& \multicolumn{5}{c}{Single lepton} \\
\hline
\multirow{4}{*}{Leptons} & \multicolumn{3}{c|}{two same-sign leptons with} & \multicolumn{2}{c}{three leptons with} \\
& \multicolumn{3}{c|}{$\pT > 27, 20$~\textrm{GeV}} & \multicolumn{2}{c}{$\pT > 27, 20, 20$~\textrm{GeV}}\\
& \multicolumn{3}{c|}{} & \multicolumn{2}{c}{at least one SFOS lepton pair} \\
& \multicolumn{5}{c}{zero additional veto leptons} \\
 
\hline
$\mll$   & \multicolumn{3}{c|}{$>$ 100~\textrm{GeV}} & \multicolumn{2}{c}{-}\\
\hline
$\mlll$   & \multicolumn{3}{c|}{-} & \multicolumn{2}{c}{$>$ 100~\textrm{GeV}}\\
\hline
$b$-jets & \multicolumn{5}{c}{zero $b$-tagged small-$R$ jets} \\
\hline
$\met$   & $>$ 80~\textrm{GeV} & $>$ 60~\textrm{GeV} & \multicolumn{3}{c}{$>$ 40~\textrm{GeV}} \\
\hline
\multirow{4}{*}{Large-$R$ jets} &
at least one large-$R$ jet with & \multicolumn{2}{c|}{zero large-$R$ jets with} & at least one large-$R$ jet with & zero large-$R$ jets with \\
 
& $\pt > 200$~\GeV,\, $\abseta < 2.0$ & \multicolumn{2}{c|}{\pt > 200~\GeV,\, \abseta < 2.0} & $\pt > 200$~\GeV,\, $\abseta < 2.0$ &  $\pt > 200$~\GeV,\, $\abseta < 2.0$ \\
 
& $50$~\GeV\ $< \mJ < 200$~\GeV\ & \multicolumn{2}{c|}{50~\GeV\ < \mJ < 200~\GeV} & $50$~\GeV\ $< \mJ < 200$~\GeV\ & $50$~\GeV\ $< \mJ < 200$~\GeV\  \\
 
& and pass 80\% $W$-tagger WP & \multicolumn{2}{c|}{} & &\\
\hline
\multirow{2}{*}{Small-$R$ jets} & - & \multicolumn{2}{c|}{at least two small-$R$ jets with} & - & at least two small-$R$ jets with  \\
& & \multicolumn{2}{c|}{$\pt > 20$~\GeV\ and $\abseta < 2.5$} & & $\pt > 20$~\GeV\ and $\abseta < 2.5$\\
\hline
$\mjj$ & - & $50$~\GeV\ $< \mjj < 110$~\GeV\ &  $>$ 200~\GeV\ & - & - \\
 
\bottomrule
\end{tabular}
}
\label{tab:selections}
\end{table}


It is not possible to reconstruct the heavy Higgs boson's mass because of the two undetected neutrinos in the final states. The `effective mass' \meff is found to be a powerful discriminant between the signal and most SM backgrounds since a high mass scale is expected for the signal, and it is thus used as the main observable to extract the signal in the statistical analysis described in Section~\ref{sec:results}. The effective mass is defined to be the scalar sum of the \met and the transverse momenta of the leptons and either the leading large-$R$ jet or the leading two small-$R$ jets for the boosted category and resolved category, respectively:
 
\begin{eqnarray*}
&& \textrm{boosted  category} \,\,\meff = \sum_{i}{\pt^{i} (\textrm{lepton})} + \pt (\textrm{leading J}) + \met, \\
&& \textrm{resolved category} \,\,\meff = \sum_{i}{\pt^{i} (\textrm{lepton})} + \pt (\textrm{leading j}) + \pt (\textrm{sub-leading j}) + \met. \\
\end{eqnarray*}


\section{Background estimation}
\label{sec:backgrounds}
 
The SM processes that mimic the $\ell^\pm\nu\ell^\pm\nu qq$ signal signature can be mainly grouped into four categories:
 
\begin{itemize}
 
\item Processes that produce at least three prompt leptons or two prompt leptons with the same electric charge. The main contributions come from $WZ$\,+\,jets (referred to as `$WZ$ background'), same-sign $WW$\,+\,jets (referred to as `ss$WW$ background'), $WWW$ (referred to as `$WWW$ background'), with other small contributions from $ttV$, $tZq$, $tth$, $WWZ$, $WZZ$, $ZZZ$ (referred to as `Other background'). These backgrounds are estimated with MC simulations, except for the backgrounds from $WZ$ and ss$WW$ production, for which the normalisations are corrected using data in dedicated CRs as defined in Section~\ref{sec:selections}.
 
\item Processes that produce two or three prompt charged leptons, but the charge of one lepton is misidentified (referred to as `charge-flip background'). A data-driven method is used to estimate this background and details are provided in Section~\ref{sec:CF}.
 
\item Processes that have one or two non-prompt leptons originating either from misidentified jets or from semileptonic decays of heavy-flavour hadrons (referred to as `non-prompt background'). A data-driven method is used to estimate this background and details are provided in Section~\ref{sec:NP}
 
\item The $W\gamma$\,+\,jets or $Z\gamma$\,+\,jets processes where the photon converts to an electron--positron pair (referred to as `photon conversion background'). A data-driven method is used to estimate this background and details are provided in Section~\ref{sec:PC}
 
\end{itemize}
 
\subsection{Electron charge-flip background}
\label{sec:CF}
The charge-flip background originates from processes that produce oppositely charged prompt leptons, where one lepton's charge is misidentified and results in final states reconstructed as having two same-sign leptons. The charge-flip background is only significant for electrons and is mainly due to interactions of the electron with material in the ID.
The dominant contributions for this background come from $\ttbar$, $W^+W^-$+\,jets, and $Z$\,+\,jets processes, and are strongly suppressed by the charge misidentification suppression BDT and the kinematic requirements on $\met$ and $\mll$.
 
The electron charge misidentification rate is measured in a data sample enriched in $Z\rightarrow e^+e^-$ events (referred to as the $Zee$ CR) selected by requiring two nominal electrons with an invariant mass between 75~\GeV\ and 105~\GeV.
Non-$Zee$ backgrounds are estimated from the total number of events in two sideband regions, defined by $60~\GeV < m_{ee} < 75$~\GeV\ and $105~\GeV < m_{ee} < 120$~\GeV. The non-$Zee$ backgrounds are then subtracted from the $Z$ mass region. The sample contains mostly opposite-charge di-electron events, with a small fraction of same-sign di-electron events. The fraction of same-sign di-electron events is used to extract the charge misidentification rate as a function of the electron $\pt$ and $\abseta$ using a likelihood fit method described in Ref.~\cite{EGAM-2018-01}, taking into account that either electron in the same-sign pair could be the misidentified one.  This rate is found to range between 0.01\% and 4\%, where higher values are obtained at large rapidities because of the larger amount of material traversed by the electrons, and at high $\pT$ because of the larger probability of an incorrect determination of the track curvature. The charge-flip background is estimated in a given region by applying the misidentification rates to data events satisfying all selection criteria except that the two electrons must be oppositely charged.
 
The statistical uncertainty of this estimate varies between 1\% and 10\%. Additional systematic uncertainties are considered by comparing the estimated nominal rate with the rate derived by: i) varying the sidebands by 4~\GeV, ii) using the $Z$\,+\,jets MC simulation directly, and iii) using MC simulation for background subtraction in the $Zee$ CR. The impact of systematic uncertainties on the charge-flip background yield is approximately 10\%, and is dominated by the uncertainty from using the $Z$\,+\,jets MC simulation directly.
 
\subsection{Non-prompt background}
\label{sec:NP}
 
The estimation of the non-prompt background assumes that these contributions can be extrapolated from a fake-lepton CR, enriched in non-prompt leptons, with a so-called fake-factor.
Events that pass the kinematic requirements of the signal regions but contain one nominal lepton and one `jet-like' lepton are selected in the fake-lepton CR. Jet-like electrons have to satisfy the likelihood-based \textit{Medium} identification~\cite{EGAM-2018-01}, while the isolation requirement is removed.
Jet-like muons have the impact parameter requirement loosened to $|d_0/\sigma_{d_0}|<10$, and the isolation requirement removed. Jet-like leptons are also required to fail at least one of the nominal lepton selections to ensure that the definitions of nominal and jet-like leptons are mutually exclusive. Simulation shows that the dominant contribution to this background stems from real muons or electrons from heavy-flavour hadrons that undergo semileptonic decays, and is heavily suppressed by the isolation and zero $b$-tagged small-$R$ jet requirements, as well as the kinematic requirements on $\mll$.
 
Events in the fake-lepton CR are scaled by fake-factors to predict the non-prompt-lepton background in the SR.
The fake-factors are calculated in control regions with selections designed to enhance the contribution from backgrounds with non-prompt leptons.
The control region selections require two same-sign leptons and exactly one $b$-tagged small-$R$ jet. One of the same-sign leptons must fulfil either the nominal criteria or those of a jet-like lepton, while the other lepton must satisfy the nominal lepton criteria.
The fake-factor is defined as the ratio of the number of events in the control region with all selected leptons fulfilling the nominal lepton criteria, to the number of events in the same region with one of the selected leptons satisfying the requirements of a jet-like lepton.
The fake-factors are calculated separately for electrons and muons as a function of the lepton $\pT$ and $\abseta$. The SM processes with prompt leptons and the charge-flip contributions are subtracted in the CR. For the electron and muon fake-factor measurements, the lepton with the second-highest $\pT$ is assumed to be the non-prompt one. This assumption is true for more than 90\% of events, based on generator-level information in the MC event record, and the potential bias can be covered by the systematic uncertainties as discussed below. The fake-factor dependency on the electron $\abseta$ is found to be negligible.
The electron fake-factors are then measured in three different electron $\pT$ bins separated by boundaries at 30~\GeV\ and 40~\GeV. The statistical uncertainty is found to be approximately 20\% in each bin.
A strong $\abseta$ dependency is found for the muon fake-factors, and their values are estimated in three $\abseta$ bins:  $0<\abseta<0.5$, $0.5<\abseta<1.5$ and $\abseta>1.5$. The statistical uncertainty is approximately 20\% in the first two bins, and approximately 30\% in the last bin. The fake-factor dependency on the muon $\pT$ was also checked and found to be negligible.
 
Apart from the statistical uncertainty, a set of systematic uncertainties is also considered for the estimation of the fake-factors as follows: i) estimating the fake-factors in the inclusive $\pt$ and $\abseta$ region; ii) varying the normalisation of the SM processes with prompt leptons and electron charge-flip background when doing the subtraction; iii) varying the $b$-tagging algorithm working points used for the $\ttbar$-enriched CR definition; iv) estimating the fake-factors with MC simulation directly in both the SRs and $\ttbar$-enriched regions, and treating the difference as a systematic uncertainty to take into account any potential fake-factor difference between SRs and $\ttbar$-enriched regions. The overall systematic uncertainty amounts to approximately 13\% (10\%) for the electron (muon) fake-factors, with the dominant contribution coming from fake-factor estimation in the inclusive $\pt$ and $\abseta$ region.

\subsection{Photon conversion background}
\label{sec:PC}
 
The photon conversion background can contribute in the SRs if the photon is misreconstructed as an electron. This background originates primarily from the $W\gamma$ process and is evaluated using a data-driven method similar to the non-prompt-lepton background estimation by introducing `photon-like' electrons. A photon-like electron is an object reconstructed like a nominal electron except that the track has no hit in the innermost layer of the pixel detector and the photon-conversion electron suppression requirements are not applied. In order to determine the photon fake-factors, a $Z\gamma$-enriched region is selected by requiring two nominal muons, no $b$-tagged small-$R$ jets, and one nominal or photon-like electron. The trilepton invariant mass must satisfy $80~\GeV < m_{\mu\mu e} < 100$~\GeV.
The photon fake-factor is defined as the ratio of the number of events in the $Z\gamma$-enriched region with the selected electron required to fulfil nominal electron criteria, to the number of events in the same region with the selected electron satisfying photon-like electron requirements. The SM processes with prompt leptons are subtracted in the $Z\gamma$-enriched region.

The photon fake-factors are measured in two electron $\pT$ bins separated by a boundary at 25~\GeV. The statistical uncertainty is found to be approximately 10\% in each bin. The fake-factor dependency on the electron $\abseta$ was also checked and found to be negligible.
A photon-conversion electron CR is then filled with events passing the signal region kinematic requirements, but containing one nominal lepton and one photon-like electron. Events in this CR are scaled by the photon fake-factor to predict the photon conversion background in the SR.
 
In a similar way to the non-prompt background, the photon conversion background fake-factor derived from the inclusive $\pt$ region is considered as one of the systematic uncertainties, together with the uncertainties from background subtraction. In addition, possible differences between $W\gamma$ and $Z\gamma$ photon fake-factors are checked with MC simulation and found to be negligible. The overall systematic uncertainty is found to be approximately 8\%.

\subsection{Validation of background estimates}
 
\begin{figure}[htbp]
\begin{center}
\begin{tabular}{cccc}
\includegraphics[width=0.45\linewidth]{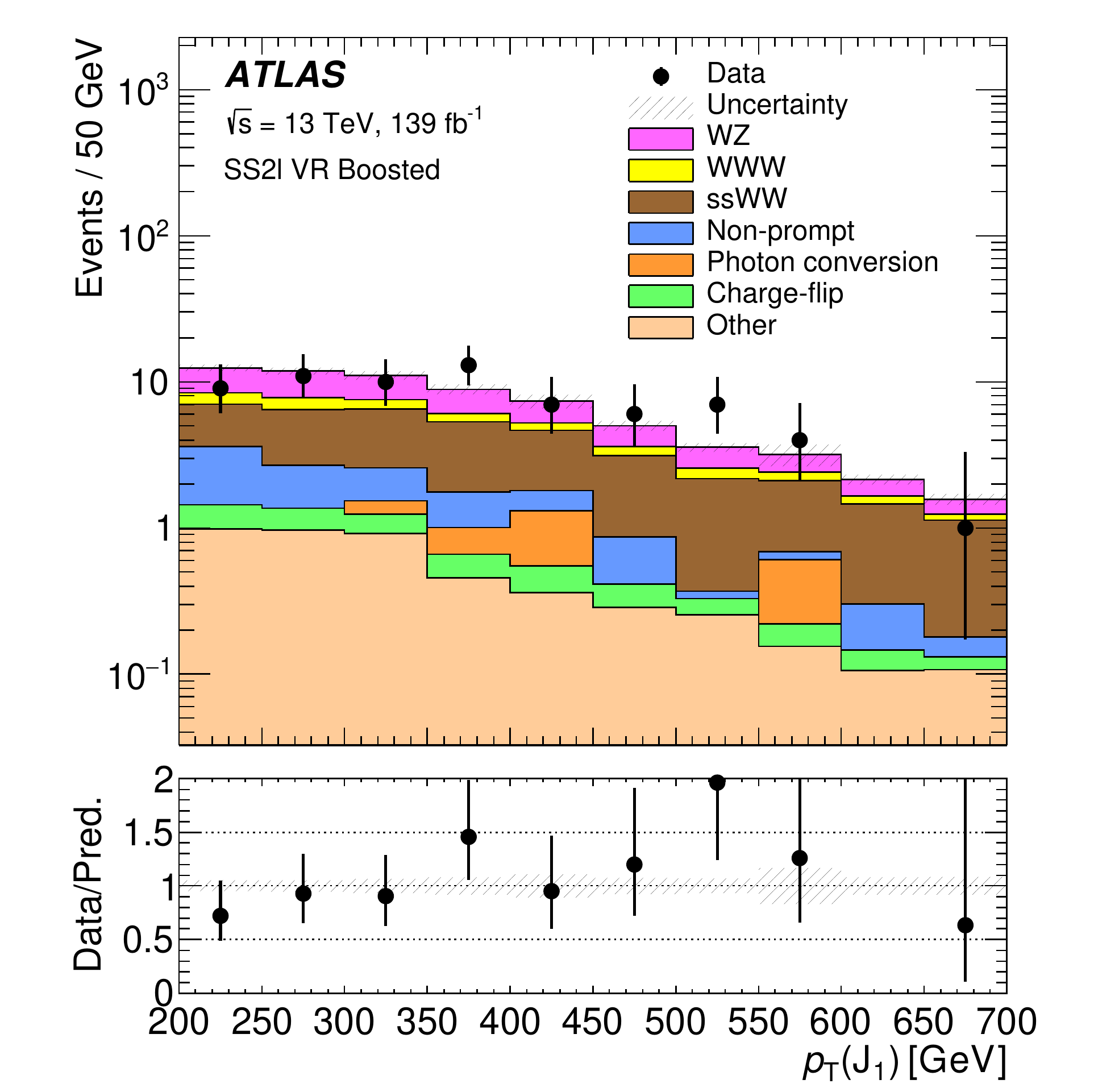}
\includegraphics[width=0.45\linewidth]{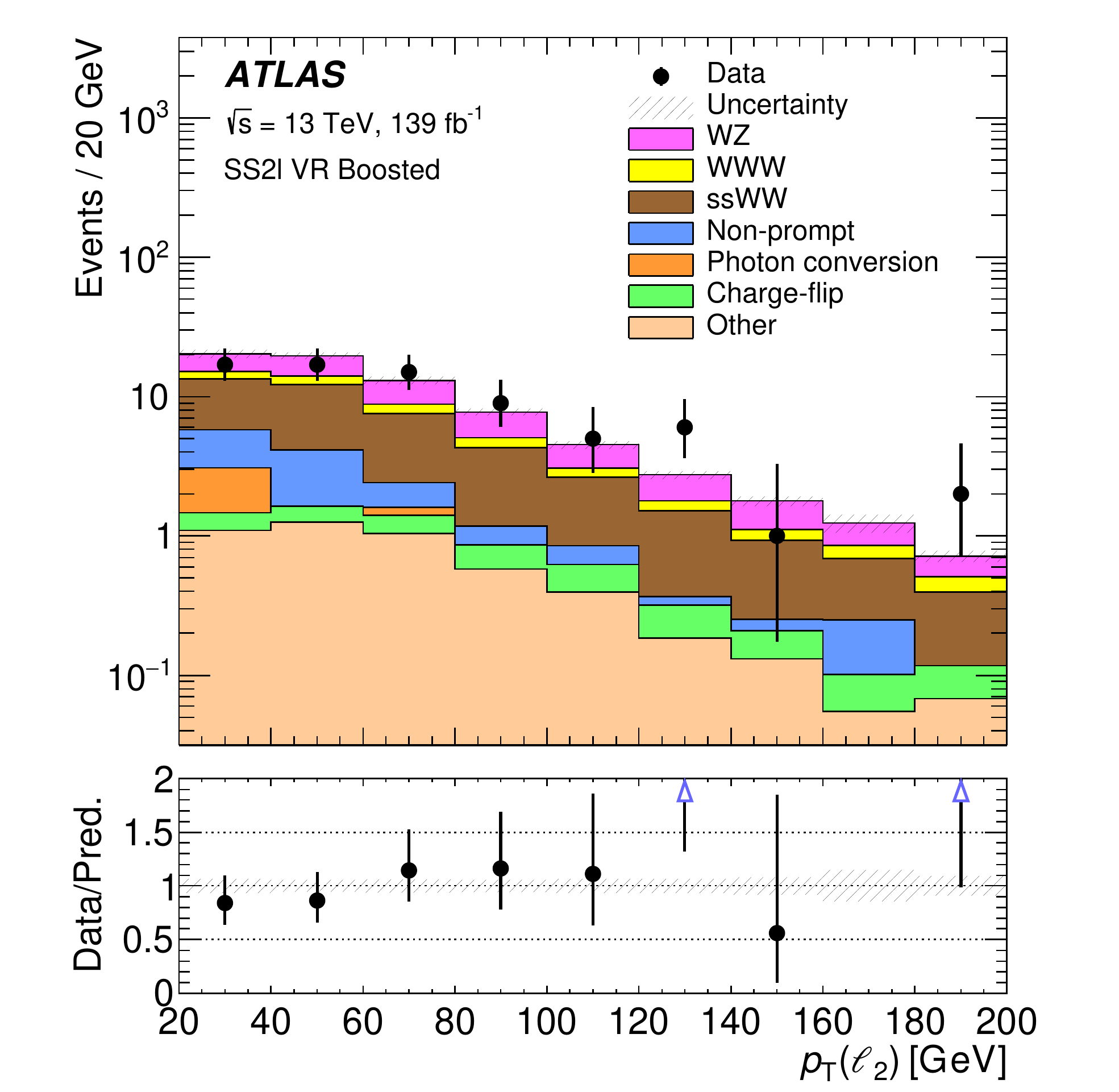} \\
\includegraphics[width=0.45\linewidth]{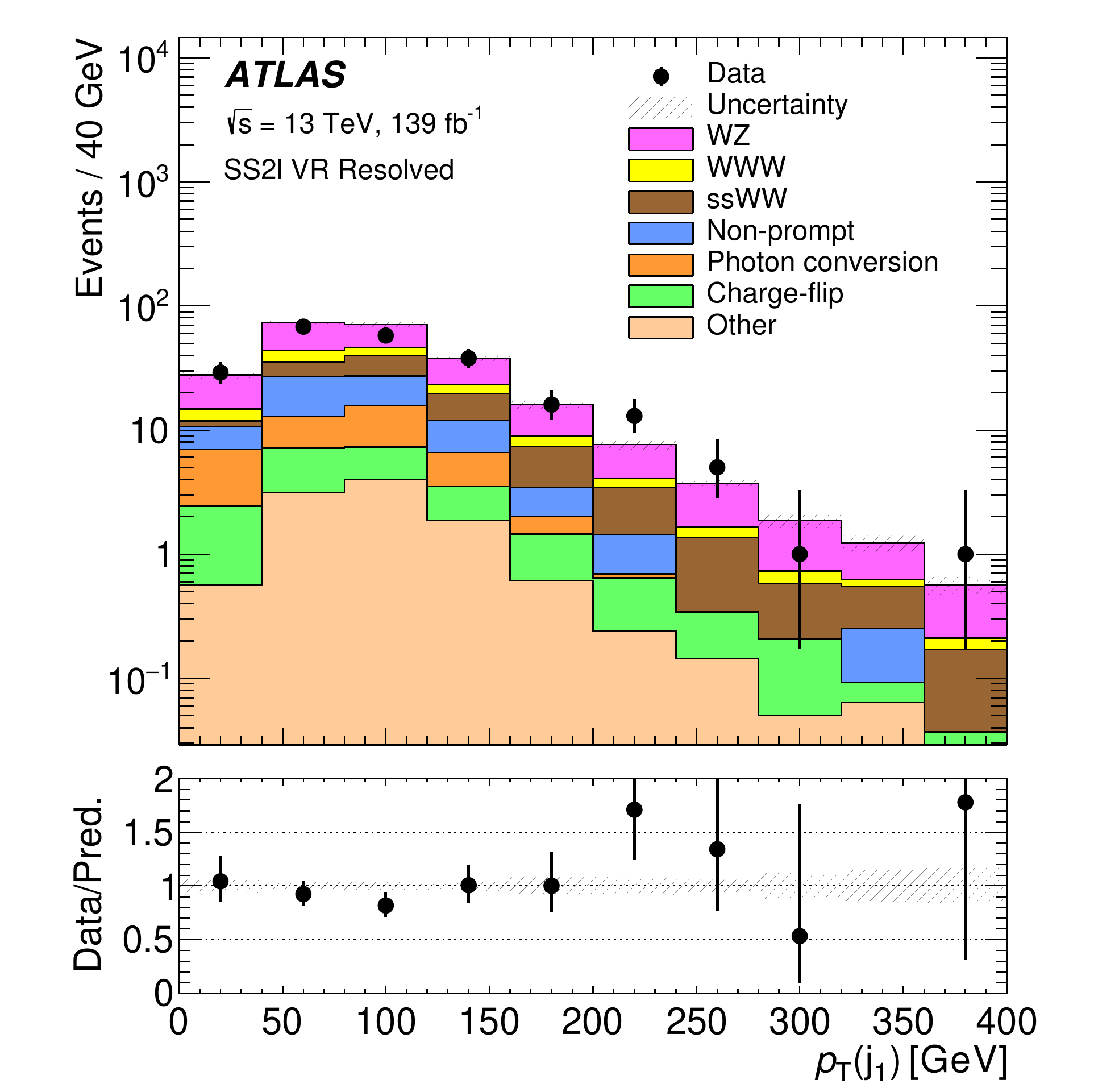}
\includegraphics[width=0.45\linewidth]{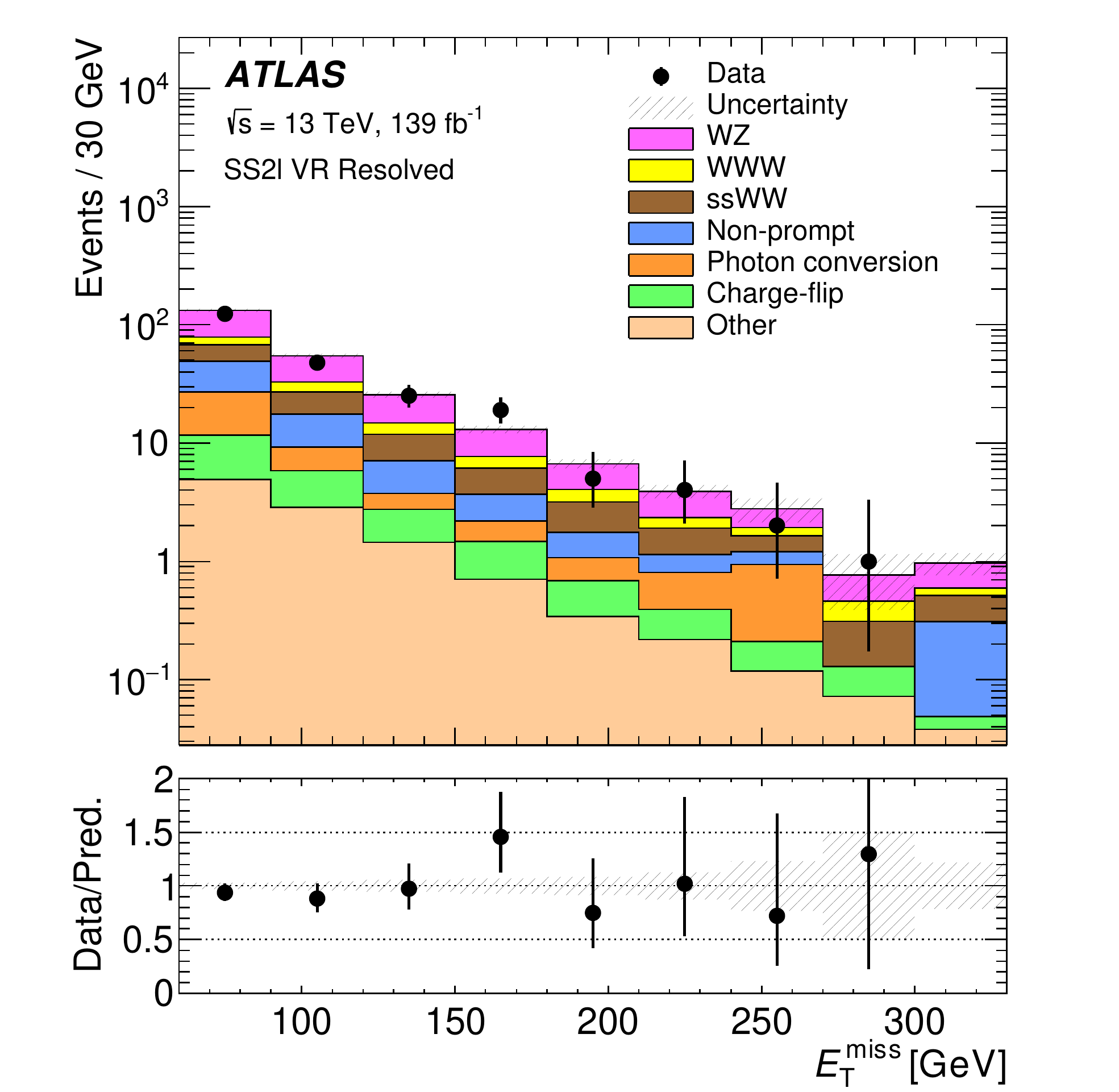} \\
\end{tabular}
\end{center}
\caption{Comparison between data and SM predictions for the large-$R$ jet $\pt$ (top left) and the sub-leading lepton $\pt$ (top right) in the boosted VR, and for the leading small-$R$ jet $\pt$ (bottom left) and $\met$ (bottom right) in the resolved VR. Predictions from simulation are scaled to the integrated luminosity of the data using the theoretical cross-section of each sample. $WZ$ and ss$WW$ backgrounds are also scaled by the normalisation factors from the global likelihood fit. The background predictions are shown as filled histograms. The size of the statistical uncertainty for the sum of the backgrounds is indicated by the hatched band. The lower panel displays the ratio of data to the total prediction. The blue triangles indicate bins where the ratio is non-zero and outside the vertical range of the plot.}
 
\label{fig:plotsVR}
\end{figure}

Two validation regions (VR) are used to test the general background predictions in the boosted and resolved categories. They are defined to be close to the signal regions, with the large-$R$ jet $W$-tagging requirement inverted in the boosted category and the $\mjj$ requirement inverted in the resolved category. Events with $\mjj > 200$~\GeV\ in the resolved category are removed in order to avoid overlap with the ss$WW$ CR. Kinematic distributions are checked and good agreement between the data and the prediction is observed in the boosted and resolved VRs, as shown in Figure~\ref{fig:plotsVR}. Data-driven methods detailed in this section are used to estimate the charge-flip, non-prompt and photon-conversion backgrounds in the VRs. Predictions from simulation are scaled to the integrated luminosity of the data using the theoretical cross-section of each sample. The $WZ$ and ss$WW$ backgrounds are also scaled by normalisation factors from the global likelihood fit as detailed in Section~\ref{sec:results}. Only the statistical uncertainty is shown in Figure~\ref{fig:plotsVR}.

\clearpage


\section{Systematic uncertainties}
\label{sec:systematics}
 
The sources of systematic uncertainty can be broadly divided into three groups: those of experimental nature and related to the detector and reconstruction performance, those of theoretical origin and associated with modelling of the simulated background and signal processes, and those related to the data-driven background estimation. The effect of the systematic uncertainties are described together with the results in Section~\ref{sec:results} and Table~\ref{tab:breakdown}, respectively.

\subsection{Experimental uncertainties}
 
Experimental uncertainties arise from the measurement of the luminosity, the modelling of pile-up in the simulation, the trigger selection, the reconstruction and identification of electrons, muons and jets, and the  $\met$ calculation.
 
The uncertainty in the combined 2015--2018 integrated luminosity is 1.7\%. It is derived following a methodology described in Ref.~\cite{ATLAS-CONF-2019-021}, and using the LUCID-detector for the baseline luminosity measurements~\cite{Avoni:2018iuv}. An uncertainty associated with the modelling of pile-up in the simulation is included to cover the difference between the predicted and measured inelastic $pp$ collision cross-sections~\cite{STDM-2015-05}.
 
Uncertainties in the reconstruction, identification, isolation and trigger efficiencies of electrons~\cite{EGAM-2018-01} and muons~\cite{MUON-2018-03} are considered, along with the uncertainty in their energy scale and resolution. These are found to have only a small impact on the result. The  lepton and jet identification efficiencies are well modelled in the simulation, and remaining differences are corrected to values measured in data. The uncertainties in the energy scale and resolution of the jets and leptons are propagated to the calculation of $\met$, which also has additional uncertainties from the modelling of the underlying event and momentum scale, momentum resolution and reconstruction efficiency of the tracks used to compute the soft-term~\cite{PERF-2016-07, ATLAS-CONF-2018-023}.
 
The uncertainties in the small-$R$ jet energy scale and resolution have contributions from in situ calibration studies, from the dependency on pile-up activity and on the flavour composition of the jets~\cite{PERF-2016-04,PERF-2011-04}.
 
Uncertainties in the efficiencies for tagging $b$-jets and for mis-tagging light-flavour jets are determined from ${t\bar{t}}$ and jet control samples, respectively~\cite{FTAG-2018-01, FTAG-2020-08, ATLAS-CONF-2018-006}.
For large-$R$ jets, the uncertainties in the energy and mass scales rely on a comparison of the ratio of calorimeter-based to track-based measurements in dijet data and simulation, as described in Ref.~\cite{JETM-2018-02}.
The efficiency of the $W$ boson tagging is estimated using data control samples, following the technique described in Ref.~\cite{ATLAS-PHYS-PUB-2020-008}. The efficiency for large-$R$ jet selection from $W$ boson decays is estimated using $\ttbar$ control samples for $\pT < 600$~\GeV. The measurement is extrapolated to the higher $\pT$ region with additional uncertainties estimated from simulations~\cite{ATL-PHYS-PUB-2020-017}. The efficiency for background large-$R$ jets from gluons or light quarks is estimated using dijet and $\gamma$\,+\,jets samples.

\subsection{Theoretical uncertainties}
Theoretical uncertainties affect the normalisations and shapes of \meff distributions of signal and background processes. They arise from sources such as our choices of event generators, PDFs, parton shower models, and underlying-event tunes. The effects of scale and PDF uncertainties are estimated by varying the renormalisation/factorisation scales and PDF sets, respectively.
The parton shower uncertainty is evaluated at generator level by comparisons of different parton showers or corresponding scales.

The normalisations of the $WZ$ background, separated into the resolved and boosted categories, as well as the ss$WW$ background normalisation, are free to float in the global likelihood fit, as detailed in Section~\ref{sec:results}. The theoretical uncertainties of these two backgrounds are not applied to the corresponding CRs, since only normalisation information is used for these CRs in the global fit. Apart from their impact on the shape of the \meff distribution, theoretical uncertainties in the $WZ$ and ss$WW$ backgrounds also impact the SR normalisations, and can be treated as uncertainties in extrapolating from high-purity CRs to the SRs.

The combined effect of the scale and PDF uncertainties, as well as the parton shower uncertainties of the $WZ$ background, is calculated by adding in quadrature the differences between the nominal \SHERPA[2.2.2] sample and its associated systematic variations, including variations of i) the renormalisation scale by factors of 0.5 and 2, ii) the factorisation scale by factors of 0.5 and 2, iii) the CKKW merging scale from 30~\GeV\ to 15~\GeV, and iv) the parton-shower/resummation scale by factors of 0.5 and 2. The total theoretical uncertainty in the $WZ$ background yield in the boosted SR (resolved SR) is found to be 29\% (20\%), and is dominated by the scale uncertainty.
 
The same approach as used for the $WZ$ background is also used to estimate the effect of scale and PDF uncertainties for the ss$WW$ background. The effect of the parton shower uncertainty on the ss$WW$ background is estimated by comparing the nominal MC sample with a sample generated with \MGNLO+ \Herwig[7]. The total theoretical uncertainty in the ss$WW$ background yield is 31\% in the boosted SR and 25\% in the resolved SR.
 
The estimation of the theoretical uncertainties in the on-shell $WWW$ background, and of the effect of PDF and scale uncertainties on the off-shell $WWW$ ($Wh \to WWW^*$) background, uses the same approach as for the $WZ$ backgrounds. The effect of the parton shower uncertainty on the off-shell $WWW$ background is estimated by comparing the nominal sample generated by \POWHEGBOX+ \PYTHIA[8] with a sample generated by \POWHEGBOX+ \Herwig[7]. The total theoretical uncertainty in the $WWW$ background yield is 16\% in the boosted SR and 8\% in the resolved SR.
 
The theoretical uncertainties of the $WZ$, ss$WW$ and $WWW$ backgrounds are decorrelated between the resolved and boosted regions in the global likelihood fit to allow for possible differences between the two regions. Given the small contributions from the processes included in the `Other' background category, only overall normalisation uncertainties are assigned.
The uncertainties vary from 10\% to 20\% based on the latest measurements of these processes~\cite{TOPQ-2018-01, TOPQ-2016-11, deFlorian:2016spz, STDM-2017-22}.
 
For the signal samples, the effects of scale and PDF uncertainties are estimated by varying the renormalisation/factorisation scales, as well as the PDF set and parameter values used for the nominal MC samples. Parton shower uncertainties are estimated by comparing the nominal samples (\MGNLO+ \PYTHIA[8]) with alternative samples using a different parton-shower generator (\MGNLO+ \Herwig[7]).
The total theoretical uncertainty in the yields from different signal samples varies between 10\% and 40\% in the SRs.
 
\subsection{Data-driven background estimation uncertainties}
 
Uncertainties in data-driven background evaluations come mainly from statistical and systematic uncertainties in the charge misidentification rate, lepton fake-factor, and photon-like electron fake-factor. More details can be found in Section~\ref{sec:backgrounds}.


\section{Results}
\label{sec:results}
 
\subsection{Statistical analysis}
The statistical analysis is based on the HistFitter framework~\cite{Baak:2014wma}. A binned likelihood function is constructed as the product of Poisson probability terms over the bins of the input distributions involving the numbers of data events and the expected signal and background yields, taking into account the effects of the floating background normalisations and the systematic uncertainties. A profile-likelihood-ratio test statistic is used to determine whether the background-only hypothesis is compatible with the observed data. The signal-plus-background hypothesis for the production of a heavy Higgs boson is tested, parameterised with the signal-strength parameter, $\mu$, defined as the ratio of the extracted cross-section to the injected hypothesised signal cross-section.
Maximum-likelihood fits to the observed binned distributions of the $\meff$ discriminants in the two SRs and to the numbers of observed events in CRs are performed simultaneously. The $\meff$ distributions are divided into 3 (5) bins for the boosted (resolved) SRs. The bins are of variable size to optimise the fit performance, while keeping the statistical uncertainty of the background contributions in each bin no larger than 10\%. The normalisations of the $WZ$ background, separated into boosted and resolved regions, as well as the normalisations of the ss$WW$ background, are free parameters in these fits and are constrained by the data in both the high-purity CRs and the SRs. The effect of systematic uncertainties in the signal and background predictions is described by nuisance parameters, which are constrained by Gaussian or log-normal probability density functions. For each nuisance parameter, the constraint is added as a penalty term to the likelihood, which decreases as soon as the nuisance parameter is shifted away from its nominal value. The statistical uncertainties of background predictions from simulation are included through one nuisance parameter per bin, using the Beeston--Barlow technique~\cite{Barlow:1993dm}.

\subsection{Data and background comparisons}
 
\begin{figure}[htbp]
\begin{center}
\begin{tabular}{cccc}
\includegraphics[width=0.45\linewidth]{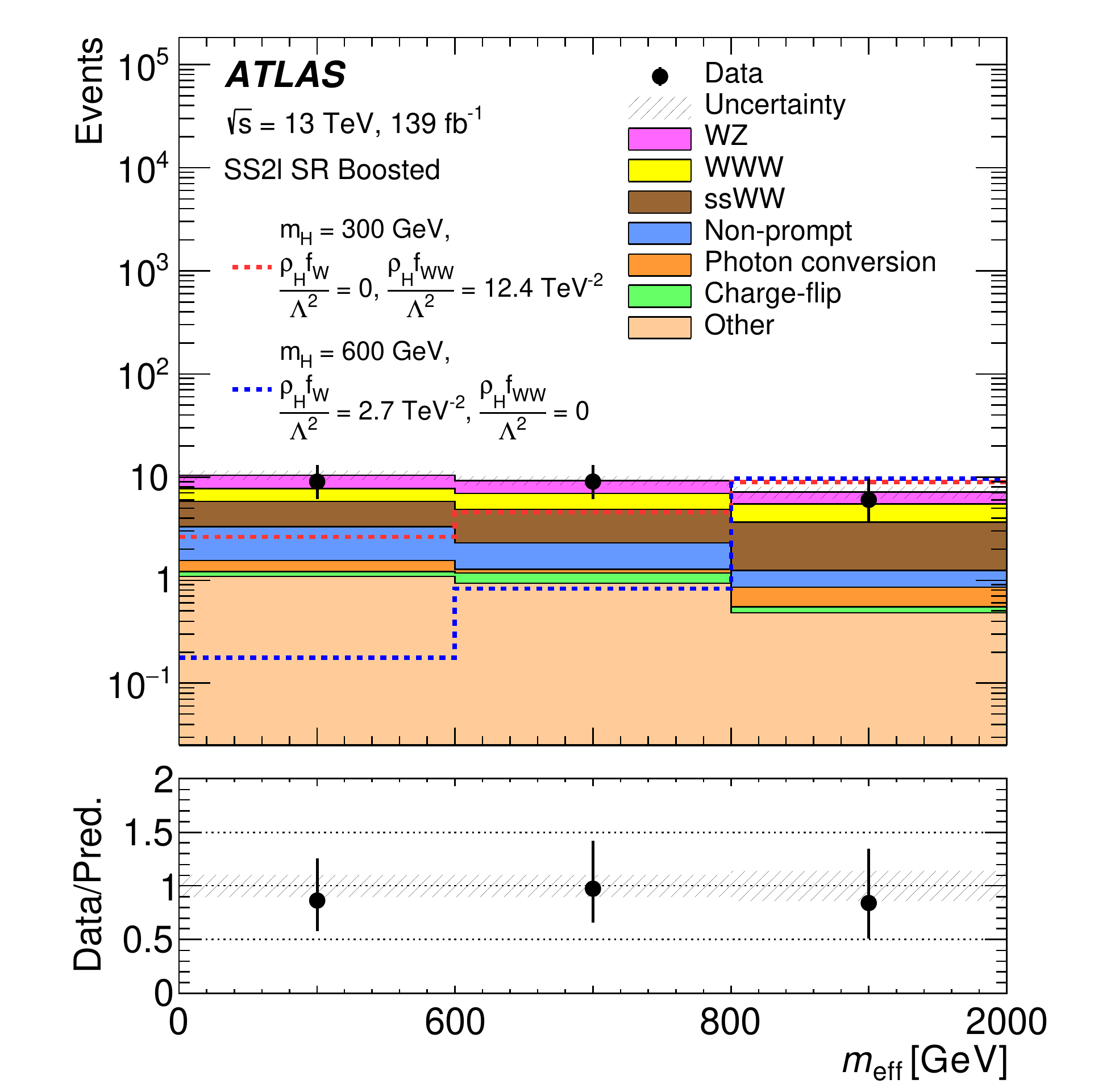}
\includegraphics[width=0.45\linewidth]{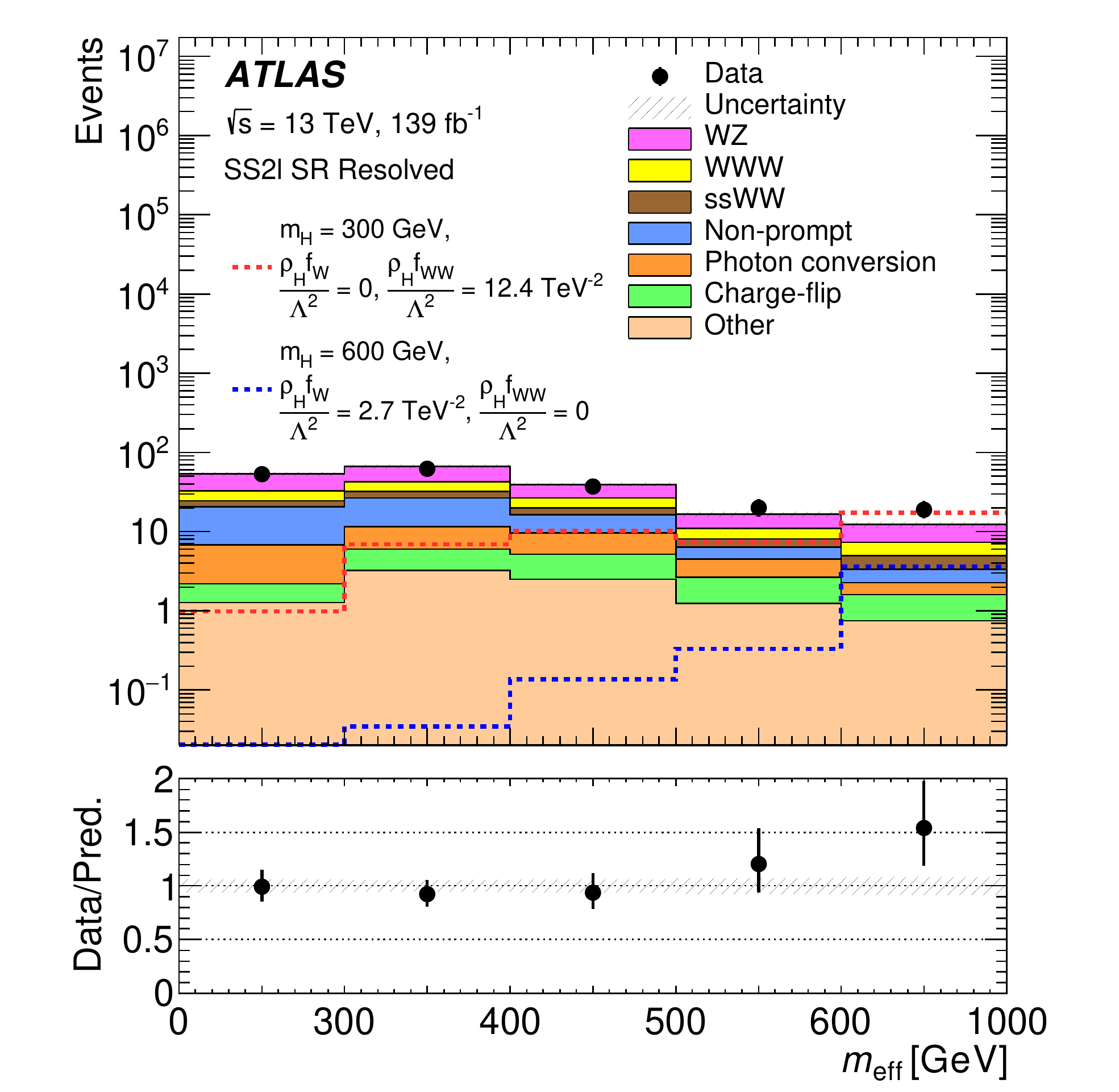} \\
\end{tabular}
\end{center}
\caption{Comparison between data and SM predictions for the $\meff$ distributions in the boosted SR (left) and the resolved SR (right). The background predictions are obtained through a background-only simultaneous fit and are shown as filled histograms. The last bin includes overflow entries. The size of the combined statistical and systematic uncertainty for the sum of the fitted backgrounds is indicated by the hatched band. The ratio of the data to the sum of the fitted backgrounds is shown in the lower panel. Two benchmark signal samples, as indicated in the legend, are also shown as unstacked unfilled histograms normalised to the integrated luminosity of the data using the theoretical cross-sections.}
\label{fig:plotsMeff}
\end{figure}

\begin{figure}[htbp]
\begin{center}
\begin{tabular}{cccc}
\includegraphics[width=0.45\linewidth]{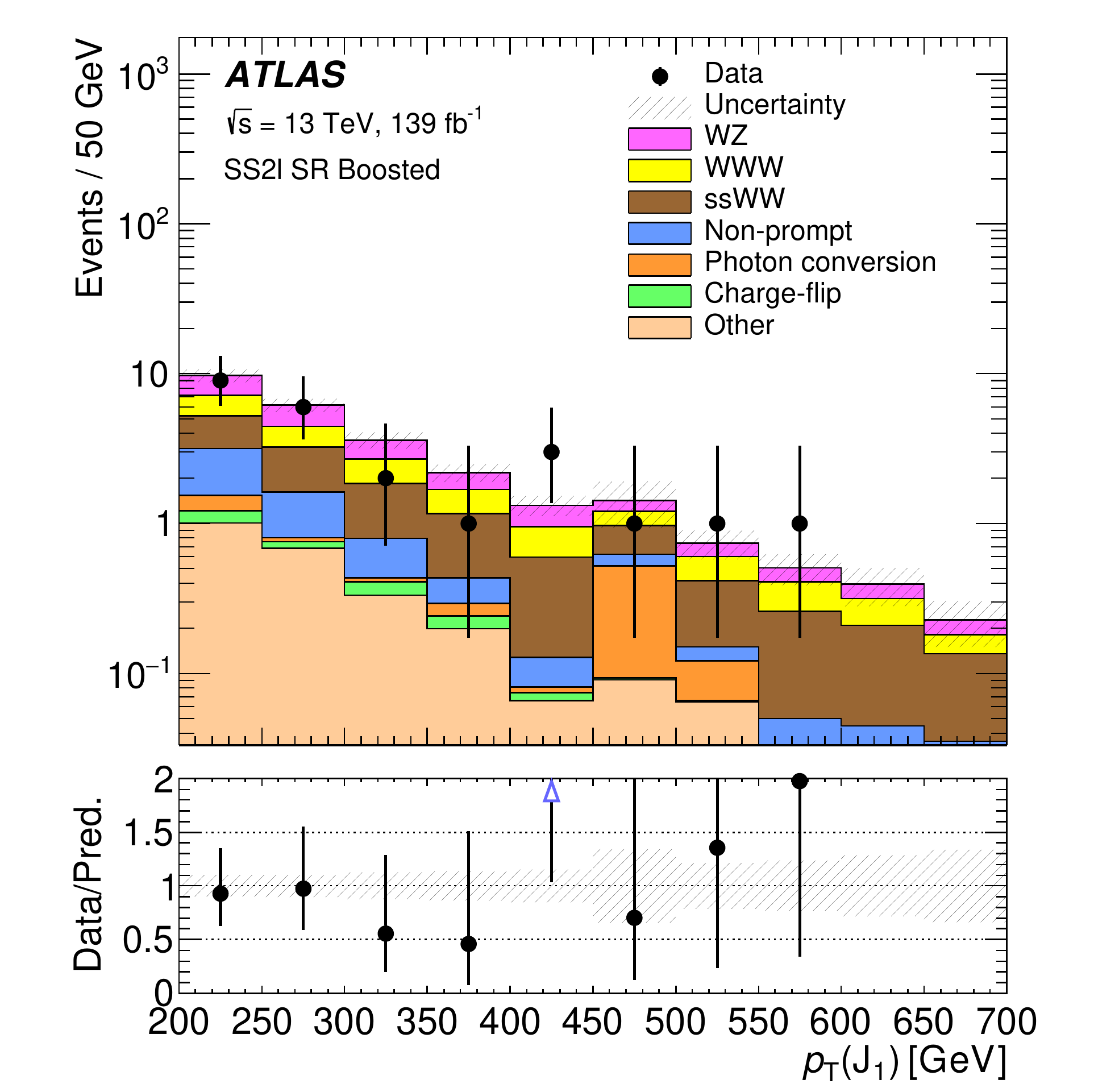}
\includegraphics[width=0.45\linewidth]{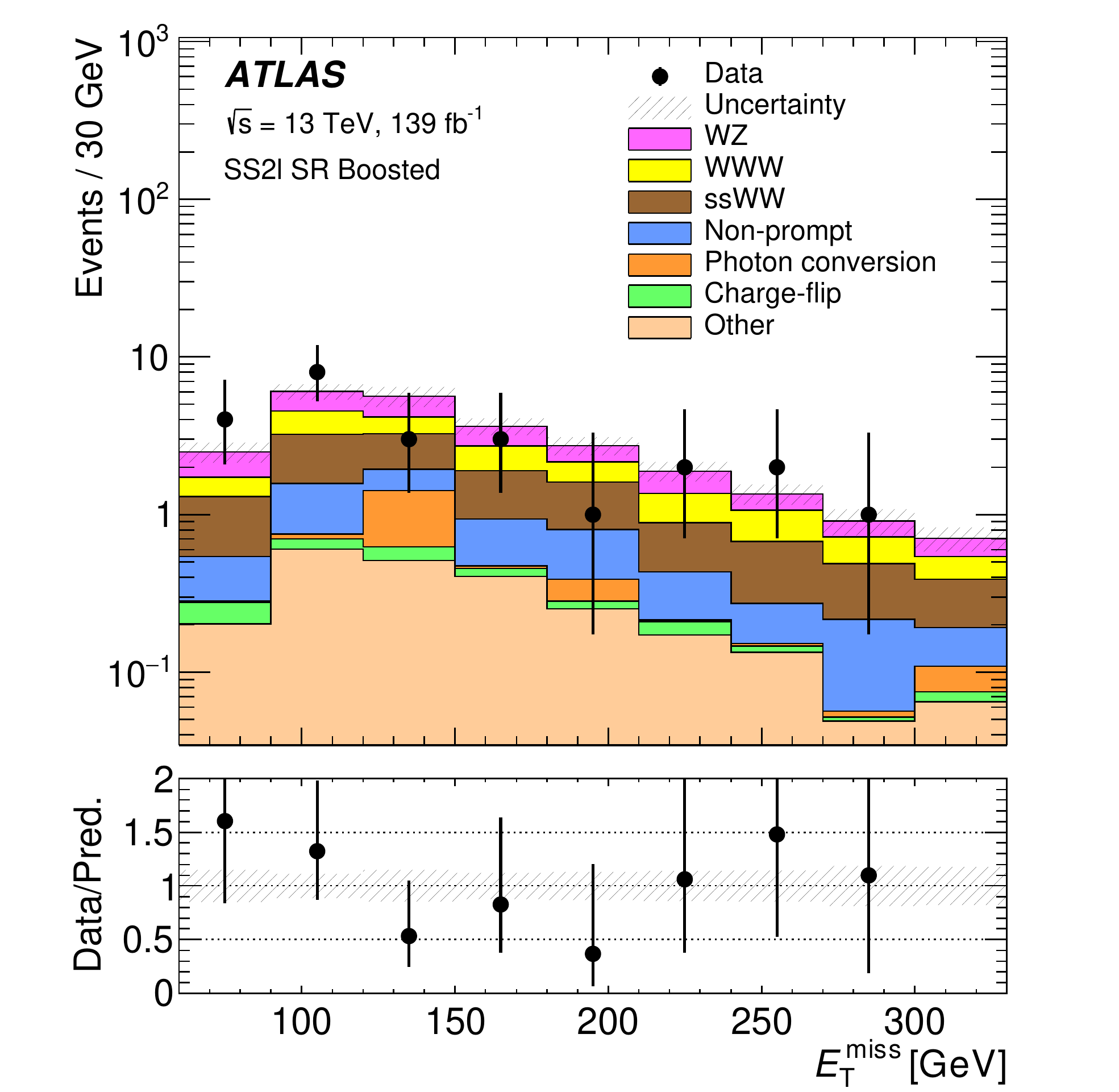}\\
\includegraphics[width=0.45\linewidth]{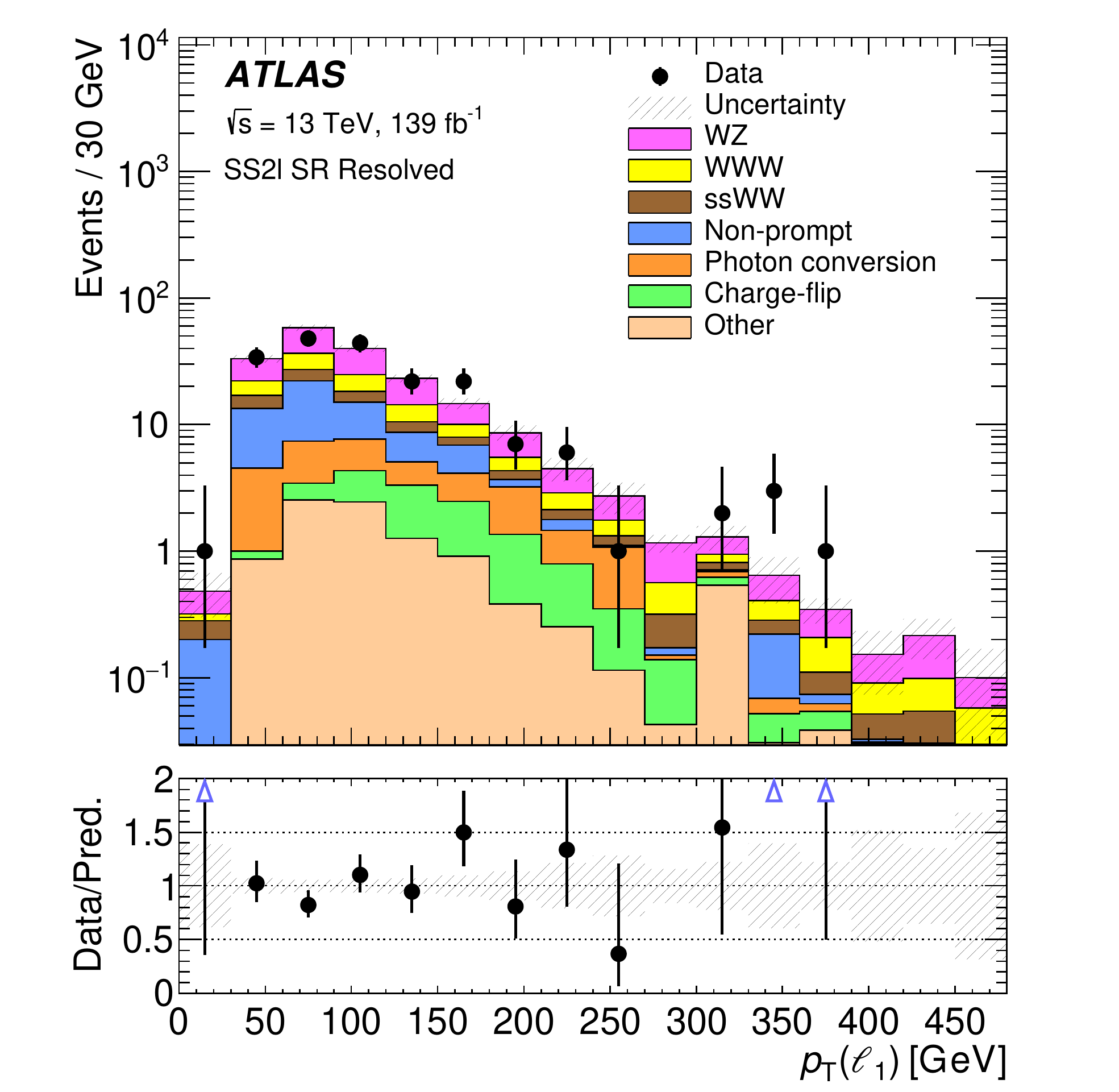}
\includegraphics[width=0.45\linewidth]{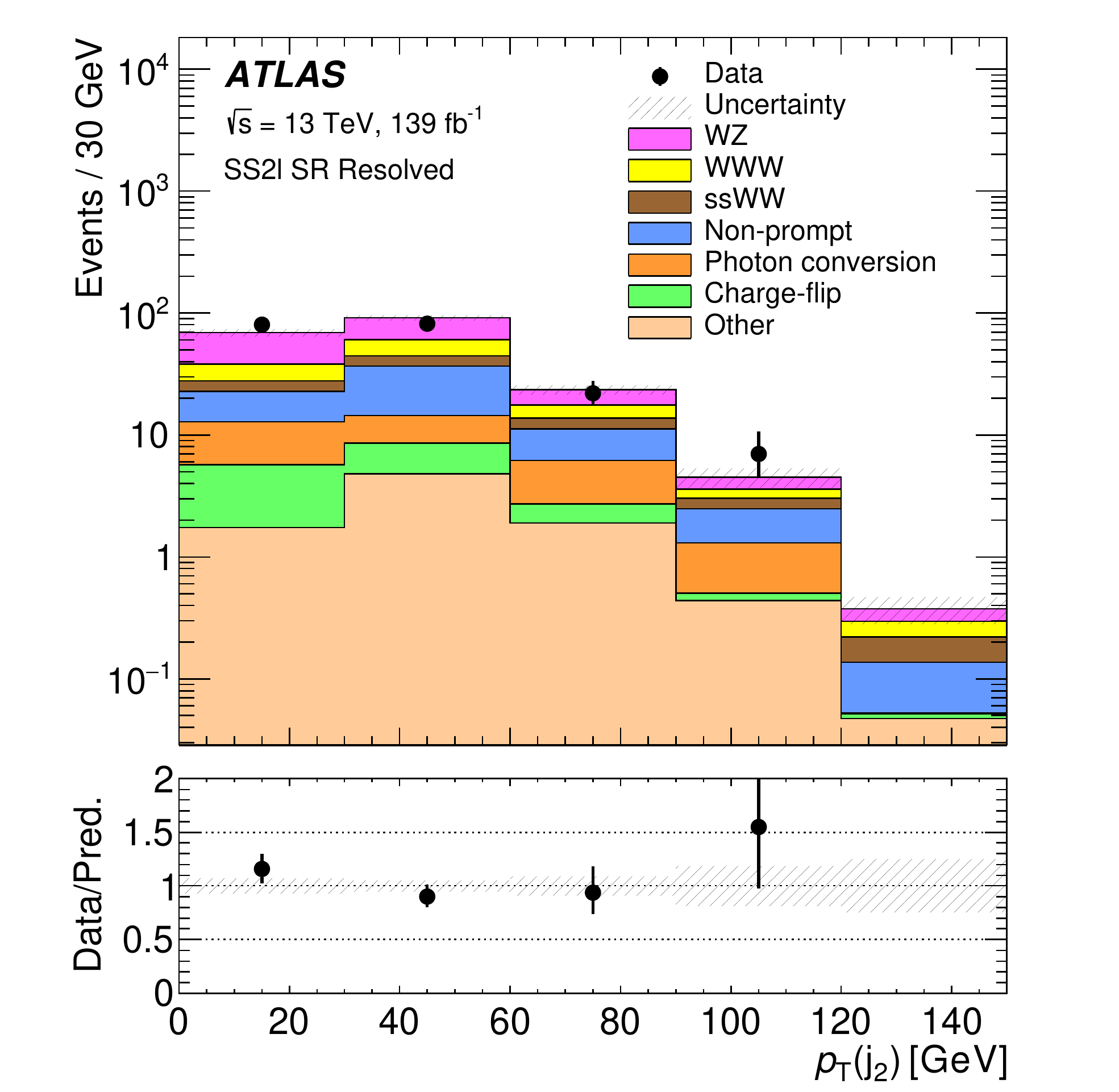} \\
\end{tabular}
\end{center}
\caption{Comparison between data and SM predictions for the large-$R$ jet $\pt$ (top left) and $\met$ (top right) in the boosted SR, and for the leading lepton $\pt$ (bottom left) and the sub-leading small-$R$ jet $\pt$ (bottom right) in the resolved SR. The background predictions are obtained through a background-only simultaneous fit and are shown as filled histograms. The size of the combined statistical and systematic uncertainty for the sum of the fitted backgrounds is indicated by the hatched band. The ratio of the data to the sum of the fitted backgrounds is shown in the lower panel. The blue triangles indicate bins where the ratio is non-zero and outside the vertical range of the plot.}
\label{fig:plotsOthers}
\end{figure}

To test the compatibility of the data and the background expectations, the data are first fit to the background-only hypothesis. Good agreement between the data and the post-fit background contributions is found for the $\meff$ distributions in the SRs and event yields in CRs. The post-fit normalisation factors of the unconstrained $WZ$ background in the boosted and resolved regions are $0.93 \pm 0.07$ and $0.83 \pm 0.03$, respectively. For the ss$WW$ background, the extracted normalisation factor is $1.54 \pm 0.18$. The errors represent the combined statistical and systematic uncertainties, but do not include theoretical uncertainties related to normalisation in the respective CRs. The extracted normalisation factors are consistent with the results from dedicated $WZ$\,+\,jets~\cite{STDM-2018-03} and ss$WW$\,+\,jets~\cite{STDM-2017-06} measurements. Table~\ref{tab:yields} shows the post-fit background event yields from different sources in all SRs and CRs, compared with the numbers of events in data.
The post-fit $\meff$ distributions in the SRs are shown in Figure~\ref{fig:plotsMeff}, where good agreement between the data and the post-fit background contributions is observed. Figure~\ref{fig:plotsOthers} shows a few selected post-fit kinematic distributions in the SRs. No significant discrepancies are observed.

\begin{table}
\caption{Background predictions and data yields for each signal region and control region. The background predictions are obtained through a background-only simultaneous fit. All systematic uncertainties are included. The individual uncertainties can be correlated, and do not necessarily add in quadrature to equal the total background uncertainty. An entry of `--' indicates that a specific background component is negligible in a certain region, or that no simulated events are left after the analysis selections.}
\resizebox{\textwidth}{!}{
\small
\begin{tabular*}{\textwidth}{@{\extracolsep{\fill}}lccccc}
\toprule
\textbf{Yields}           		& Boosted SR		& Resolved SR            	& Boosted $WZ$ CR		& Resolved $WZ$ CR	& $ssWW$ CR			\\
\midrule
Observed events		& $24$			& $191$				& $236$				& $2094$	& $567$                   					\\
\midrule
Fitted bkg events		& $26.8 \pm 2.7$~~	& $189.0 \pm 7.8$~~		& $235 \pm 15$		& $2095 \pm 46$~~		& $566 \pm 24$~		\\
\midrule
 
$WWW$			& $5.8 \pm 1.0$	& $30.4 \pm 2.9$		& ~~$1.30 \pm 0.31$		& $11.2 \pm 2.1$		& $28.5\pm 5.5$		\\
ss$WW$			& $7.5 \pm 2.3$ 	& $16.5 \pm 1.9$		& $-$          			& $-$				& $254 \pm 27$		\\
$WZ$			& $6.71 \pm 0.76$	& $68.7 \pm 5.0$		& $221 \pm 15$		& $1956 \pm 50$~~		& $150.6 \pm 5.7$~~		\\
 
Non-prompt		& $3.20 \pm 0.36$	& $39.6 \pm 6.3$		& $-$          			& $-$				& $48.6 \pm 8.8$              \\
Charge-flip		& $0.43 \pm 0.03$	& ~~$8.61 \pm 0.57$		& $-$          			& $-$          			& $22.8 \pm 1.3$              \\
Photon conversion	& $0.73 \pm 0.07$	& $17.2 \pm 1.7$		& $-$          			& $-$          			& $46.7 \pm 4.7$              \\
Other             		& $2.50 \pm 0.45$	& ~~$9.0 \pm 1.5$		& $12.3 \pm 1.6$		& $130 \pm 20$		& $14.3 \pm 2.0$              \\
\bottomrule
\end{tabular*}
}
 
\label{tab:yields}
 
\end{table}


\begin{figure}[htbp]
\begin{center}
\begin{tabular}{cccc}
\includegraphics[width=0.45\linewidth]{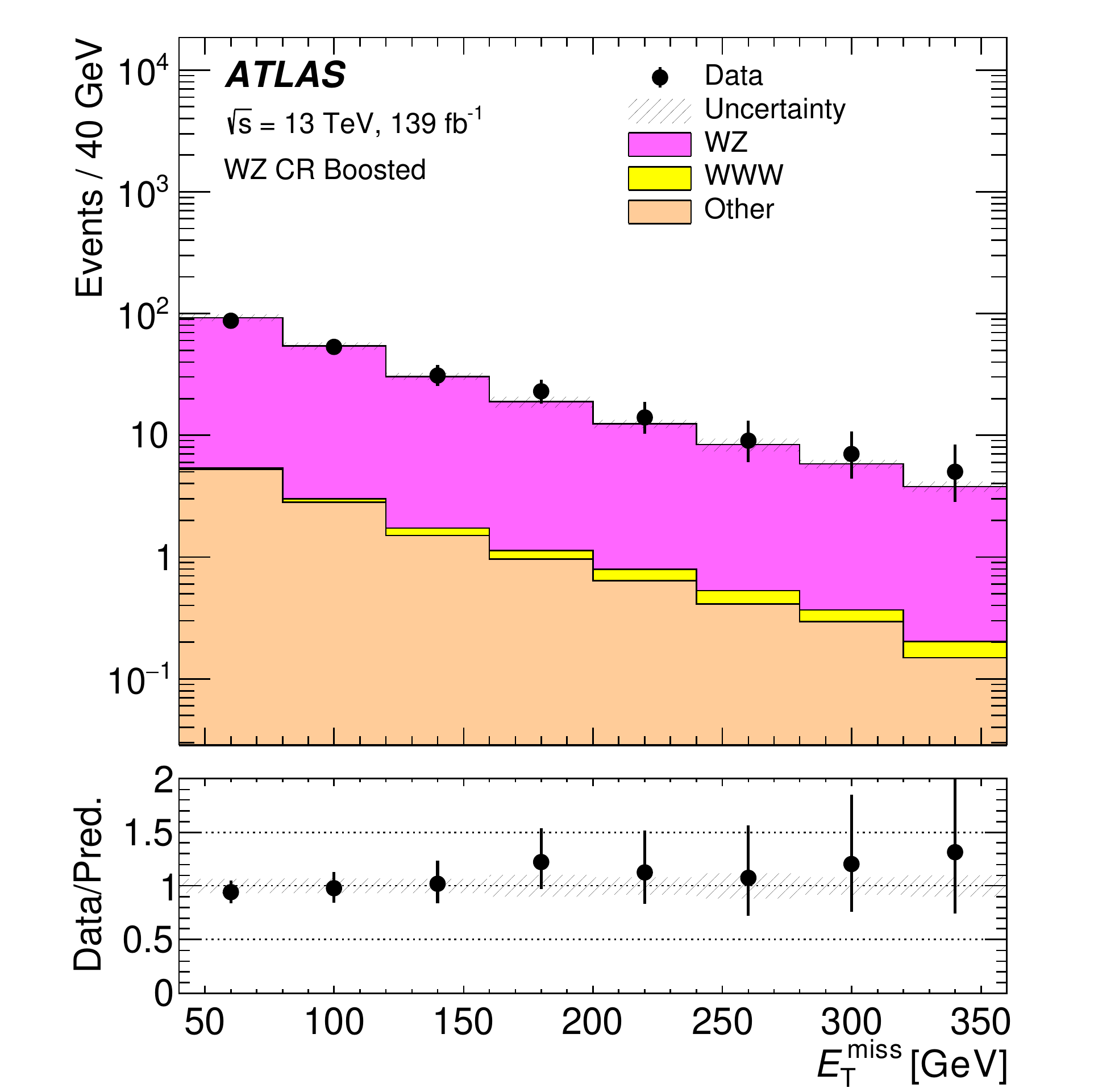}
\includegraphics[width=0.45\linewidth]{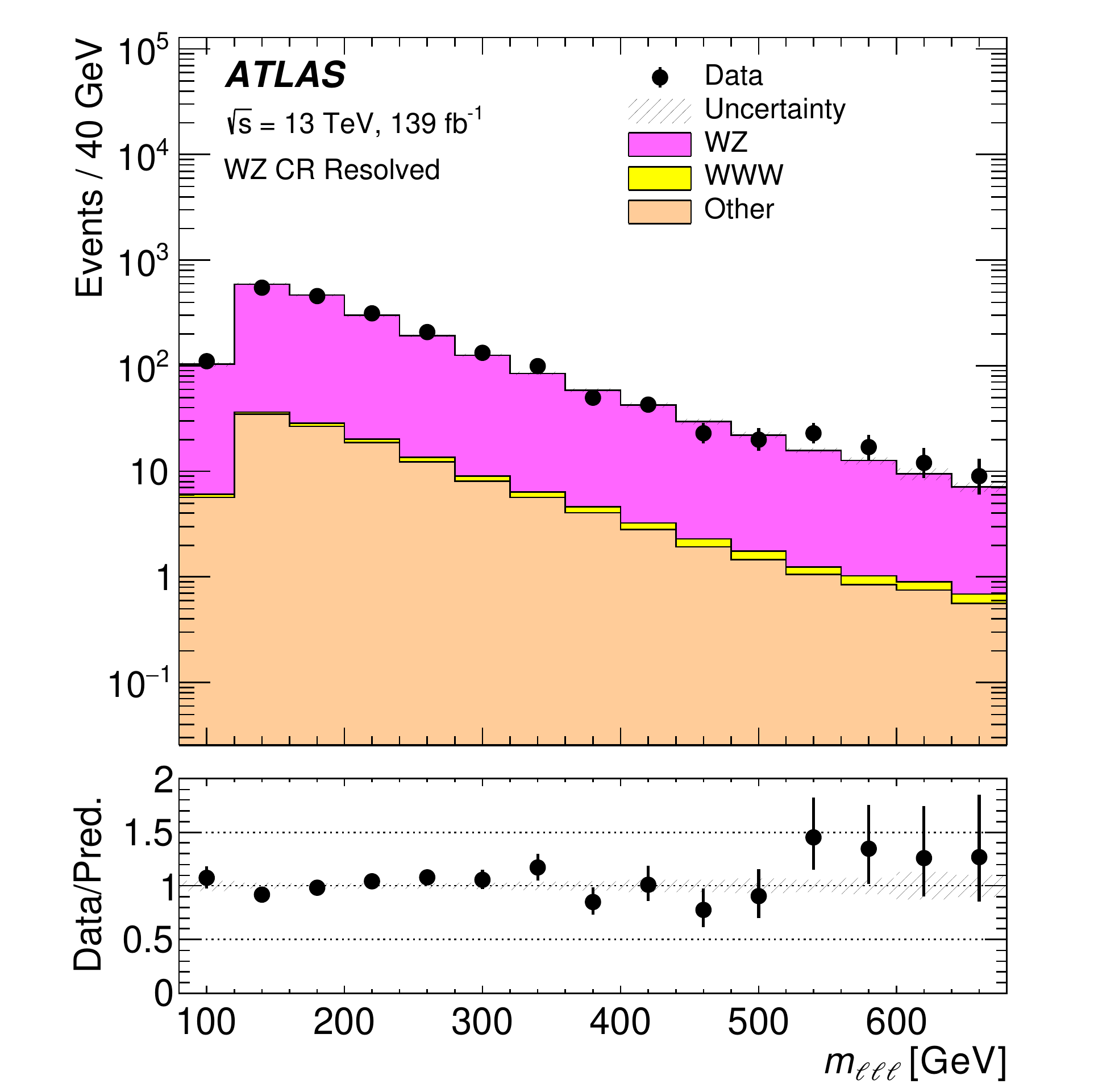} \\
\includegraphics[width=0.45\linewidth]{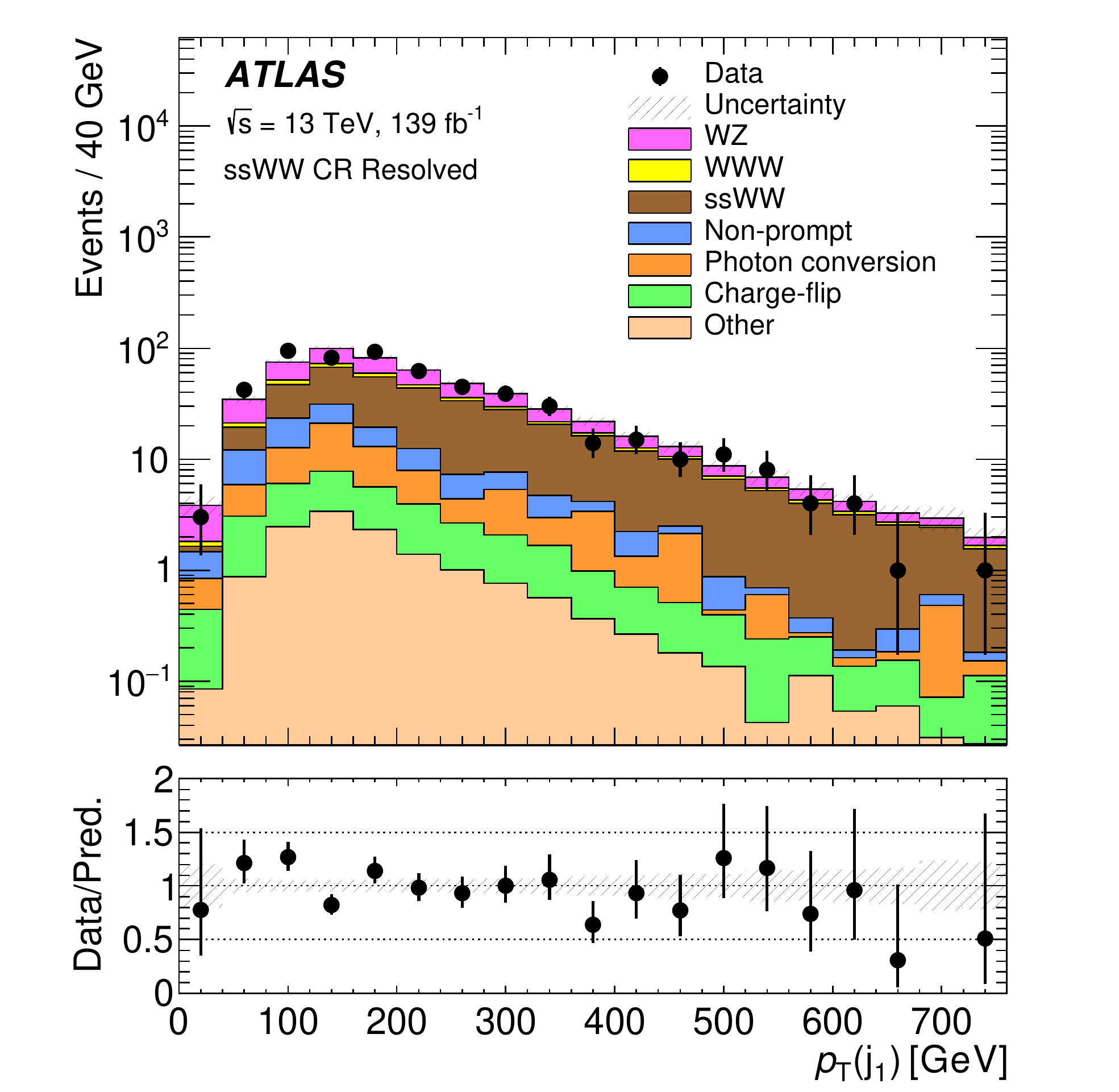}
\end{tabular}
\end{center}
\caption{Comparison between data and SM predictions for $\met$ (top left) in the boosted $WZ$ CR, for the invariant mass of the three-lepton system (top right) in the resolved $WZ$ CR, and for the leading small-$R$ jet $\pt$ (bottom) in the ss$WW$ CR. The background predictions are obtained through a background-only simultaneous fit and are shown as filled histograms. The size of the combined statistical and systematic uncertainty for the sum of the fitted backgrounds is indicated by the hatched band. The ratio of the data to the sum of the fitted backgrounds is shown in the lower panel.}
\label{fig:plotsCR}
\end{figure}

Good overall normalisation agreement between data and post-fit background contributions in CRs is seen in Table~\ref{tab:yields}, and a few selected post-fit kinematic distributions in the CRs are shown in Figure~\ref{fig:plotsCR}.
 
Table~\ref{tab:breakdown} summarises the systematic uncertainties in the background expectation for each SR. The individual sources of systematic uncertainty detailed in Section~\ref{sec:systematics} are combined into categories.
In the resolved SR, the largest uncertainty comes from the data-driven background estimation, followed by theoretical uncertainties in background modelling, the uncertainty due to the limited size of the simulated samples, and the small-$R$ jet uncertainty. In the boosted SR, the systematic uncertainties associated with the theoretical modelling of the background and with the $W$-tagger play a dominant role, followed by sizeable effects from the limited size of the simulated samples and the large-$R$ jet uncertainty.

\begin{table}
\centering
\small
\caption{
Breakdown of the dominant systematic uncertainties in background estimates in both the boosted and resolved signal regions. The background predictions are obtained through a background-only simultaneous fit.
The individual uncertainties can be correlated, and do not necessarily add in quadrature to equal
the total background uncertainty. The percentages show the size of the uncertainty relative to the total expected background. An entry of `--' indicates that a specific uncertainty component is not relevant in a certain region.}
\label{tab:breakdown}
\begin{tabular*}{\textwidth}{@{\extracolsep{\fill}}lcc}
\toprule
\textbf{Uncertainty of channel}                                    & Boosted SR            & Resolved SR            \\
\midrule
Total systematic uncertainties & 10.0\%~~        & 4.1\% \\
\midrule
Data driven non-prompt         & 1.3\%          & 3.3\%       \\
Theoretical uncertainties         & 8.9\%           & 2.6\%        \\
MC statistical uncertainties    & 3.0\% & 1.9\% \\
Floating normalisations         & 3.5\% & 1.2\% \\
Data driven photon conversion    & 0.2\% & 0.9\% \\
$\met$         & 0.2\% & 0.7\% \\
$b$-tagging         & 0.8\% & 0.5\% \\
Data driven charge-flip         & 0.1\% & 0.3\% \\
Electron         & 0.5\% & 0.2\% \\
Muon               & 0.6\% & 0.2\% \\
Pile-up reweighting & 0.2\% & 0.2\% \\
Large-$R$ jet     & 1.1\% & 0.2\% \\
$W$-tagger         & 3.7\% & -- \\
Small-$R$ jet         & -- & 1.1\% \\
\bottomrule
\end{tabular*}
\end{table}


\FloatBarrier

\subsection{Limits on the production of heavy Higgs bosons}
 
Constraints on the production of heavy Higgs bosons are derived by repeating the fit to the signal-plus-background hypothesis. Upper limits on the production cross-sections of heavy Higgs bosons are calculated with a modified frequentist method \cite{Read:2002hq}, known as CL$_\mathrm{s}$, using the $\tilde{q}_{\mu}$ test statistic in the asymptotic approximation~\cite{Cowan:2010js}.
 
\begin{figure}[htbp]
\begin{center}
\begin{tabular}{cccc}
\includegraphics[width=0.48\linewidth]{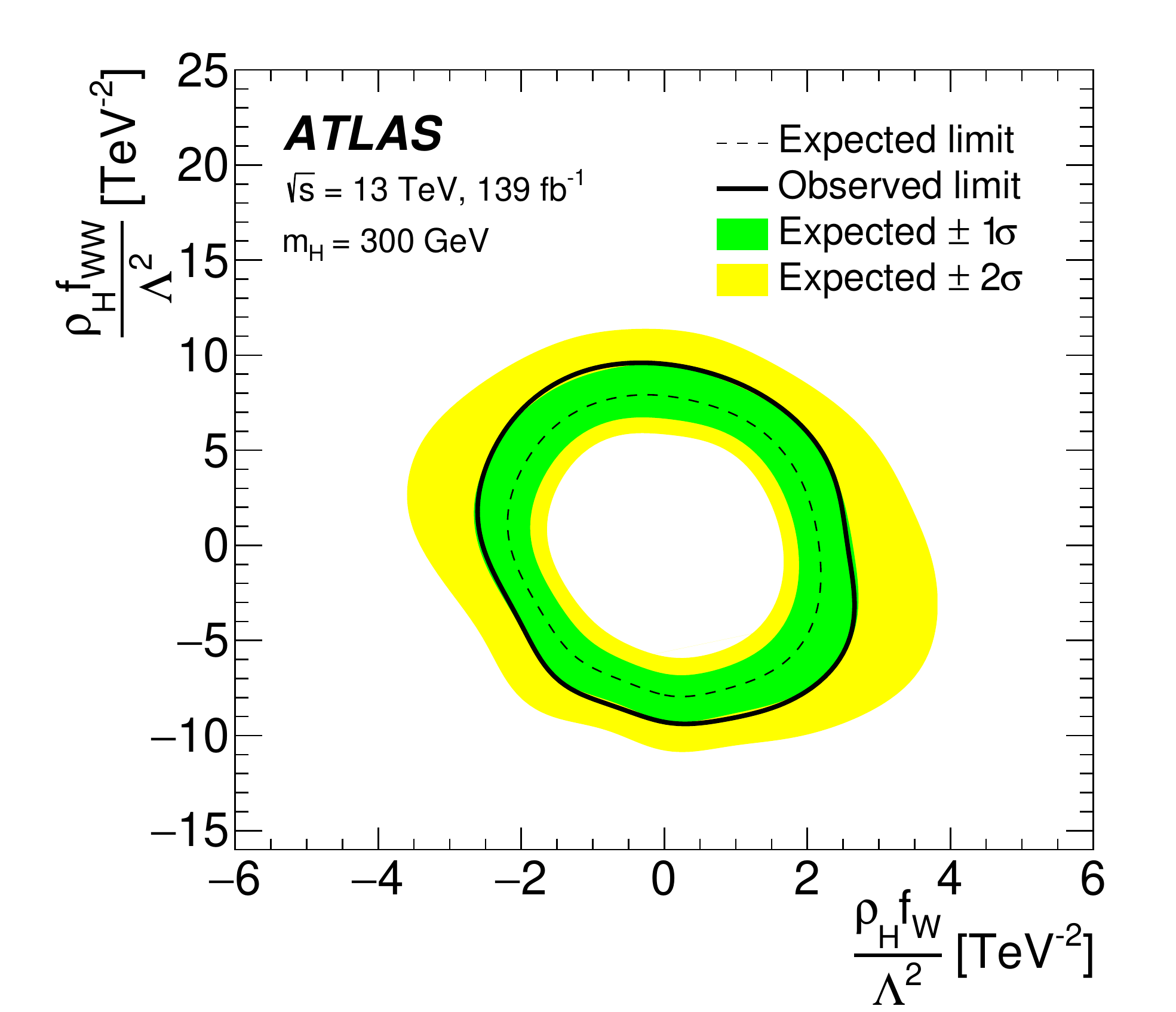}
\includegraphics[width=0.48\linewidth]{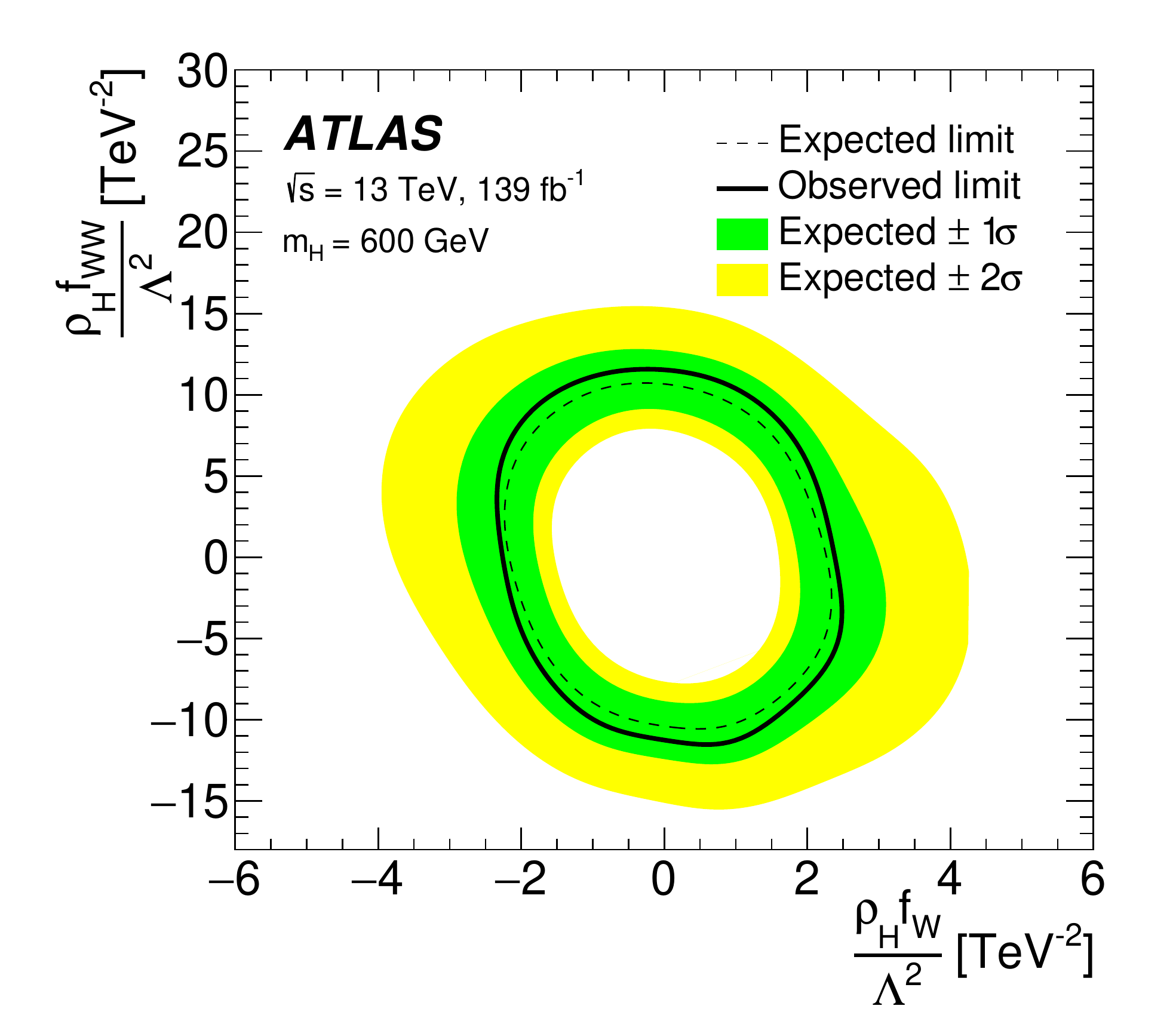}\\
\includegraphics[width=0.48\linewidth]{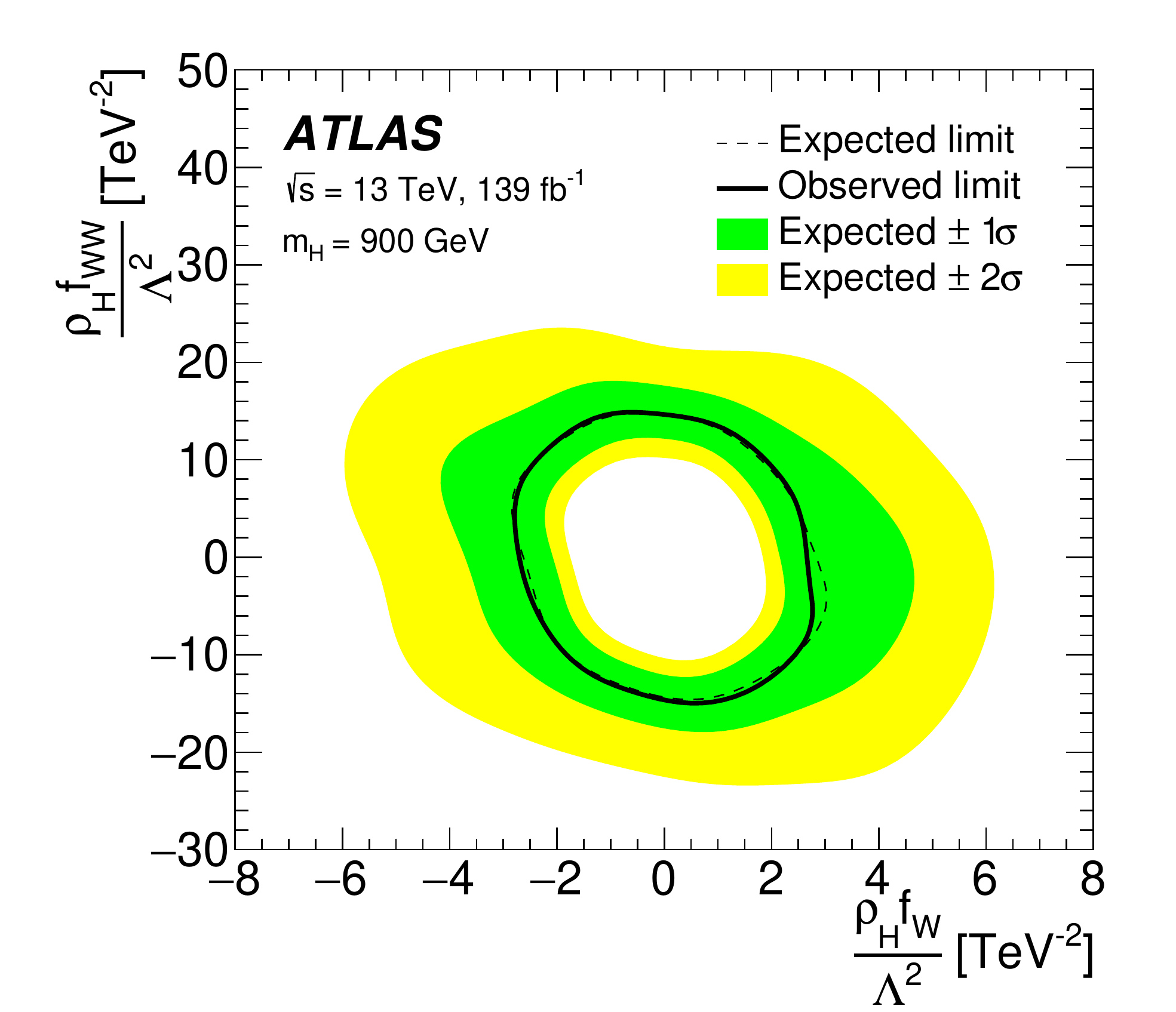}
\end{tabular}
\end{center}
\caption{Observed (black solid curve) and expected (black dashed curve) 95\% CL upper limits on the production of a heavy Higgs boson as a function of $\rho_{H}f_{W}/\Lambda^{2}$ and $\rho_{H}f_{WW}/\Lambda^{2}$ for a mass of 300~\GeV~(top left), 600~\GeV~(top right) and 900~\GeV~(bottom). The green (inner) and yellow (outer) bands represent the $\pm$1$\sigma$ and $\pm$2$\sigma$ uncertainties of the expected limits.}
\label{fig:limitfwfww}
\end{figure}

Figure~\ref{fig:limitfwfww} shows the expected and observed exclusion contours at the 95\% confidence level (CL) for signals from heavy Higgs bosons with masses of 300~\GeV, 600~\GeV\ and 900~\GeV\ as a function of the coupling strengths $\rho_{H}f_{W}/\Lambda^{2}$ and $\rho_{H}f_{WW}/\Lambda^{2}$. The scaling factor $\rho_{H}$ is set to $0.05$ and the scale $\Lambda$ is set to $5~\TeV$ as mentioned in Section~\ref{sec:phenomenology}. The hypotheses are tested for each mass value in each of the 16 radial directions of the ($f_{W}$, $f_{WW}$) space. The local $p_0$-value for the observation to be compatible with the background-only hypothesis reaches its smallest value at 300~\GeV\ with ($\rho_{H}f_{W}/\Lambda^{2}$, $\rho_{H}f_{WW}/\Lambda^{2}$) = (0, $4.9~\TeV^{-2}$), corresponding to 1.3~standard deviation. For a heavy Higgs boson with a mass of 300~\GeV, $|\rho_{H}f_{W}/\Lambda^{2}| > 2.7~\TeV^{-2}$ and $|\rho_{H}f_{WW}/\Lambda^{2}| > 10~\TeV^{-2}$ can be excluded at 95\% CL. Couplings of $|\rho_{H}f_{W}/\Lambda^{2}| > 2.5~\TeV^{-2}$ and $|\rho_{H}f_{WW}/\Lambda^{2}| > 12~\TeV^{-2}$ can be excluded for the production of a heavy Higgs boson with a mass of 600~\GeV. Similarly, for a heavy Higgs boson with a mass of 900~\GeV, $|\rho_{H}f_{W}/\Lambda^{2}| > 2.9~\TeV^{-2}$ and $|\rho_{H}f_{WW}/\Lambda^{2}| > 15~\TeV^{-2}$ can be excluded.
 
The overall excess, at the level of approximately $1\sigma$, observed for a 300~\GeV\ heavy Higgs boson is mostly due to the small excess observed in data in the rightmost bin of the resolved SR's $\meff$ distribution, as shown in Figure~\ref{fig:plotsMeff}. This is because the resolved SR dominates the sensitivity to lower-mass heavy Higgs bosons.

\begin{figure}[htbp]
\begin{center}
\begin{tabular}{cccc}
\includegraphics[width=0.48\linewidth]{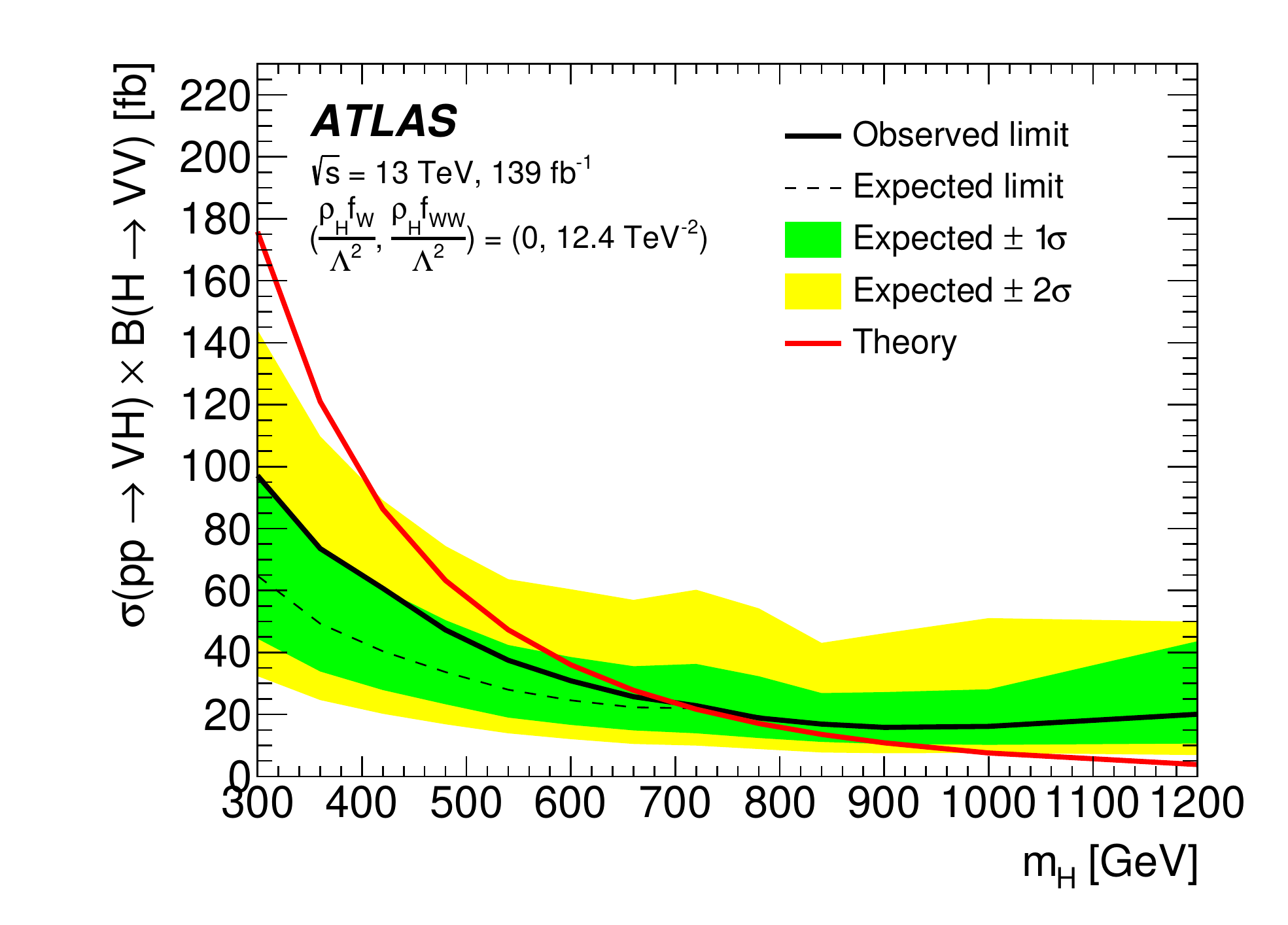}
\includegraphics[width=0.48\linewidth]{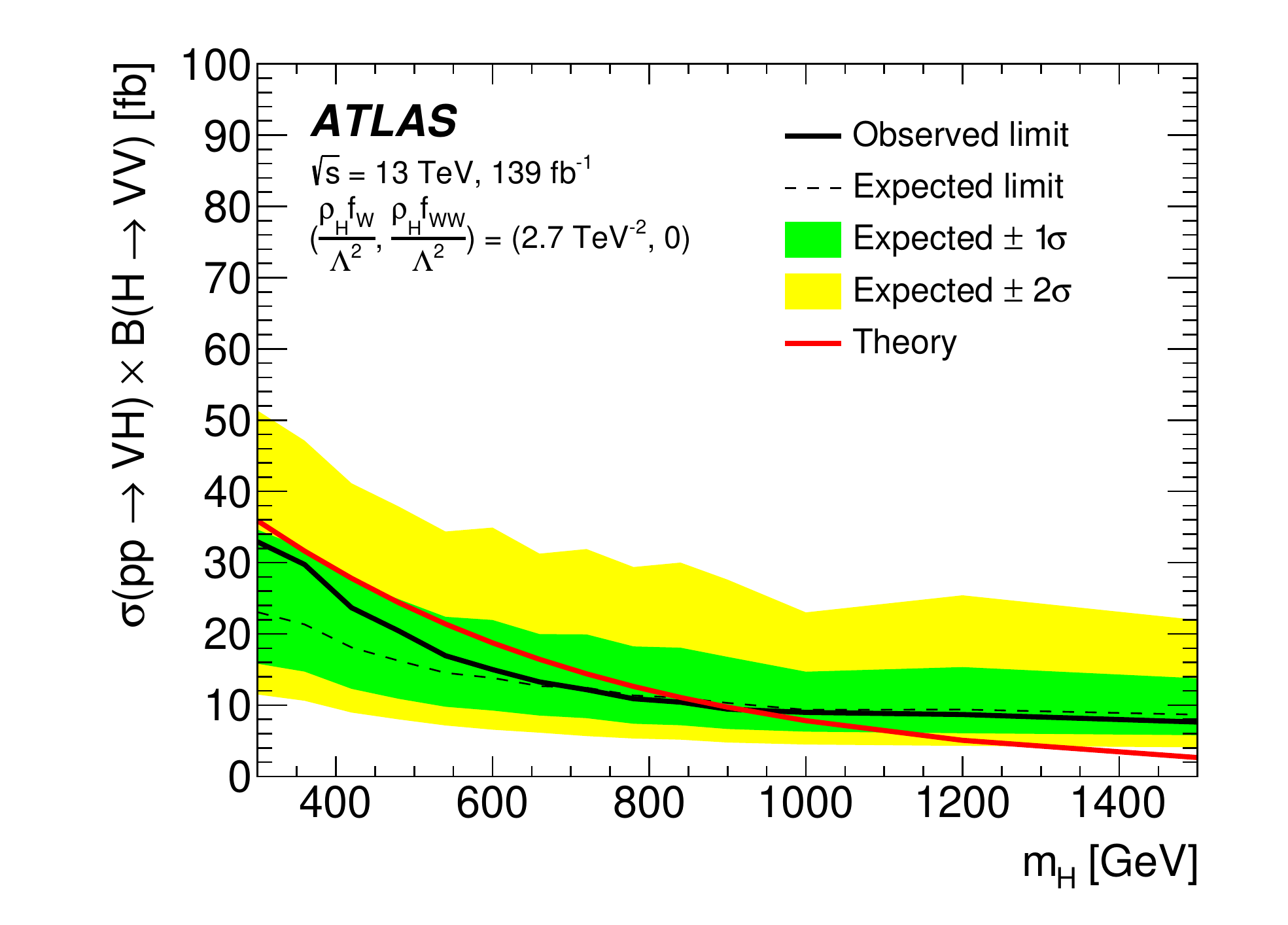} \\
\end{tabular}
\end{center}
\caption{Observed (black solid curve) and expected (black dashed curve) 95\% CL upper limits on the production cross-section times decay branching fraction of a heavy Higgs boson as a function of its mass with $(\rho_{H}f_{W}/\Lambda^{2}, \rho_{H}f_{WW}/\Lambda^{2})$ fixed to $(0,12.4~\TeV^{-2})$ (left) and $(2.7~\TeV^{-2},0)$ (right). The green (inner) and yellow (outer) bands represent the $\pm$1$\sigma$ and $\pm$2$\sigma$ uncertainties of the expected limits. The unevenness in the expected limits reflects the variations in the estimated systematic uncertainties. Theoretical predictions (red solid curve) as a function of the heavy Higgs boson's mass are overlaid.}
\label{fig:limitmass}
\end{figure}

From the ellipse of expected limits in Figure~\ref{fig:limitfwfww}, two sets of couplings $(\rho_{H}f_{W}/\Lambda^{2}, \rho_{H}f_{WW}/\Lambda^{2})$ with values $(0, 12.4~\TeV^{-2})$ and $(2.7~\TeV^{-2}, 0)$ are chosen as benchmark examples with which to explore the dependence of the results on the heavy Higgs boson's mass.
Coupling combinations on this ellipse are expected to have similar phenomenology. Although the two points chosen on the ellipse are somewhat arbitrary, they are representative. Figure~\ref{fig:limitmass} shows the expected and observed 95\% CL upper limits on the heavy Higgs boson's production cross-section as a function of its mass for those two sets of anomalous couplings.
The mass hypotheses are tested at 60~\GeV\ steps between 300 and 900~\GeV, and three additional mass points at 1000~\GeV, 1200~\GeV\ and 1500~\GeV.
For both of sets of couplings, the local $p_0$-value is smallest at 300~\GeV, corresponding to 1.3 and 0.9~standard deviations, respectively.
For the coupling choice $(\rho_{H}f_{W}/\Lambda^{2}, \rho_{H}f_{WW}/\Lambda^{2}) = (0, 12.4~\TeV^{-2})$, a heavy Higgs boson with mass up to 700~\GeV\ can be excluded, while for the choice $(2.7~\TeV^{-2}, 0)$, the range of excluded masses extends to 900~\GeV.
 
\clearpage


\section{Summary}
 
This paper presents a search for heavy Higgs bosons produced in association with a $W$ boson and decaying into a pair of $W$ bosons. The search uses proton--proton collisions at centre-of-mass energy of 13 TeV corresponding to an integrated luminosity of $139$~\fbi. The data were recorded by the ATLAS experiment between 2015 and 2018 at the LHC. The search is performed in the final states with two leptons of the same electric charge, missing transverse momentum and jets. The $\Wqq$ decay is reconstructed from two resolved small-$R$ jets or one boosted large-$R$ jet, and two corresponding signal regions are defined.
The data are found to be in good agreement with the estimated backgrounds.
 
Upper limits on the production of heavy Higgs bosons are derived as a function of the heavy Higgs boson's mass and coupling strengths to vector bosons.
For heavy Higgs bosons with masses ranging from 300~\GeV\  to 900~\GeV, the 95\% CL upper limits on the coupling strengths $|\rho_{H}f_{W}/\Lambda^{2}|$ and $|\rho_{H}f_{WW}/\Lambda^{2}|$ are in the range 2.5--2.9~\TeV$^{-2}$ and 10--15~\TeV$^{-2}$, respectively.
The most stringent exclusion ranges for the coupling strengths, $|\rho_{H}f_{W}/\Lambda^{2}| > 2.5~\TeV^{-2}$ and $|\rho_{H}f_{WW}/\Lambda^{2}| > 10~\TeV^{-2}$, are set for the production of heavy Higgs bosons with a mass of 600~\GeV\ or 300~\GeV, respectively.
The scaling factor $\rho_{H}$ is set to $0.05$ and $\Lambda$ is set to $5~\text{TeV}$ in the analysis. Heavy Higgs bosons are excluded at 95\% CL for masses up to 700~\GeV\ or 900~\GeV\ with anomalous couplings to vector bosons $(\rho_{H}f_{W}/\Lambda^{2}, \rho_{H}f_{WW}/\Lambda^{2})$ fixed at $(0,12.4~\TeV^{-2})$ or $(2.7~\TeV^{-2},0)$, respectively.


\ifconf
\else
\section*{Acknowledgements}


We thank CERN for the very successful operation of the LHC, as well as the
support staff from our institutions without whom ATLAS could not be
operated efficiently.
 
We acknowledge the support of
ANPCyT, Argentina;
YerPhI, Armenia;
ARC, Australia;
BMWFW and FWF, Austria;
ANAS, Azerbaijan;
CNPq and FAPESP, Brazil;
NSERC, NRC and CFI, Canada;
CERN;
ANID, Chile;
CAS, MOST and NSFC, China;
Minciencias, Colombia;
MEYS CR, Czech Republic;
DNRF and DNSRC, Denmark;
IN2P3-CNRS and CEA-DRF/IRFU, France;
SRNSFG, Georgia;
BMBF, HGF and MPG, Germany;
GSRI, Greece;
RGC and Hong Kong SAR, China;
ISF and Benoziyo Center, Israel;
INFN, Italy;
MEXT and JSPS, Japan;
CNRST, Morocco;
NWO, Netherlands;
RCN, Norway;
MEiN, Poland;
FCT, Portugal;
MNE/IFA, Romania;
MESTD, Serbia;
MSSR, Slovakia;
ARRS and MIZ\v{S}, Slovenia;
DSI/NRF, South Africa;
MICINN, Spain;
SRC and Wallenberg Foundation, Sweden;
SERI, SNSF and Cantons of Bern and Geneva, Switzerland;
MOST, Taiwan;
TENMAK, T\"urkiye;
STFC, United Kingdom;
DOE and NSF, United States of America.
In addition, individual groups and members have received support from
BCKDF, CANARIE, Compute Canada and CRC, Canada;
PRIMUS 21/SCI/017 and UNCE SCI/013, Czech Republic;
COST, ERC, ERDF, Horizon 2020 and Marie Sk{\l}odowska-Curie Actions, European Union;
Investissements d'Avenir Labex, Investissements d'Avenir Idex and ANR, France;
DFG and AvH Foundation, Germany;
Herakleitos, Thales and Aristeia programmes co-financed by EU-ESF and the Greek NSRF, Greece;
BSF-NSF and MINERVA, Israel;
Norwegian Financial Mechanism 2014-2021, Norway;
NCN and NAWA, Poland;
La Caixa Banking Foundation, CERCA Programme Generalitat de Catalunya and PROMETEO and GenT Programmes Generalitat Valenciana, Spain;
G\"{o}ran Gustafssons Stiftelse, Sweden;
The Royal Society and Leverhulme Trust, United Kingdom.
 
The crucial computing support from all WLCG partners is acknowledged gratefully, in particular from CERN, the ATLAS Tier-1 facilities at TRIUMF (Canada), NDGF (Denmark, Norway, Sweden), CC-IN2P3 (France), KIT/GridKA (Germany), INFN-CNAF (Italy), NL-T1 (Netherlands), PIC (Spain), ASGC (Taiwan), RAL (UK) and BNL (USA), the Tier-2 facilities worldwide and large non-WLCG resource providers. Major contributors of computing resources are listed in Ref.~\cite{ATL-SOFT-PUB-2021-003}.


\fi


\clearpage

\printbibliography
 
\clearpage
 
\begin{flushleft}
\hypersetup{urlcolor=black}
{\Large The ATLAS Collaboration}

\bigskip

\AtlasOrcid[0000-0002-6665-4934]{G.~Aad}$^\textrm{\scriptsize 101}$,
\AtlasOrcid[0000-0002-5888-2734]{B.~Abbott}$^\textrm{\scriptsize 119}$,
\AtlasOrcid[0000-0002-7248-3203]{D.C.~Abbott}$^\textrm{\scriptsize 102}$,
\AtlasOrcid[0000-0002-1002-1652]{K.~Abeling}$^\textrm{\scriptsize 55}$,
\AtlasOrcid[0000-0002-8496-9294]{S.H.~Abidi}$^\textrm{\scriptsize 29}$,
\AtlasOrcid[0000-0002-9987-2292]{A.~Aboulhorma}$^\textrm{\scriptsize 35e}$,
\AtlasOrcid[0000-0001-5329-6640]{H.~Abramowicz}$^\textrm{\scriptsize 150}$,
\AtlasOrcid[0000-0002-1599-2896]{H.~Abreu}$^\textrm{\scriptsize 149}$,
\AtlasOrcid[0000-0003-0403-3697]{Y.~Abulaiti}$^\textrm{\scriptsize 116}$,
\AtlasOrcid[0000-0003-0762-7204]{A.C.~Abusleme~Hoffman}$^\textrm{\scriptsize 136a}$
% ...
\bigskip
\\

% ...
$^{29}$Physics Department, Brookhaven National Laboratory, Upton NY; United States of America.\\
% ...
$^{35}$$^{(a)}$Facult\'e des Sciences Ain Chock, R\'eseau Universitaire de Physique des Hautes Energies - Universit\'e Hassan II, Casablanca;$^{(b)}$Facult\'{e} des Sciences, Universit\'{e} Ibn-Tofail, K\'{e}nitra;$^{(c)}$Facult\'e des Sciences Semlalia, Universit\'e Cadi Ayyad, LPHEA-Marrakech;$^{(d)}$LPMR, Facult\'e des Sciences, Universit\'e Mohamed Premier, Oujda;$^{(e)}$Facult\'e des sciences, Universit\'e Mohammed V, Rabat;$^{(f)}$Institute of Applied Physics, Mohammed VI Polytechnic University, Ben Guerir; Morocco.\\
% ...
$^{55}$II. Physikalisches Institut, Georg-August-Universit\"{a}t G\"ottingen, G\"ottingen; Germany.\\
% ...
$^{101}$CPPM, Aix-Marseille Universit\'e, CNRS/IN2P3, Marseille; France.\\
$^{102}$Department of Physics, University of Massachusetts, Amherst MA; United States of America.\\
% ...
$^{116}$Department of Physics, New York University, New York NY; United States of America.\\
% ...
$^{119}$Homer L. Dodge Department of Physics and Astronomy, University of Oklahoma, Norman OK; United States of America.\\
% ...
$^{136}$$^{(a)}$Departamento de F\'isica, Pontificia Universidad Cat\'olica de Chile, Santiago;$^{(b)}$Millennium Institute for Subatomic physics at high energy frontier (SAPHIR), Santiago;$^{(c)}$Instituto de Investigaci\'on Multidisciplinario en Ciencia y Tecnolog\'ia, y Departamento de F\'isica, Universidad de La Serena;$^{(d)}$Universidad Andres Bello, Department of Physics, Santiago;$^{(e)}$Instituto de Alta Investigaci\'on, Universidad de Tarapac\'a, Arica;$^{(f)}$Departamento de F\'isica, Universidad T\'ecnica Federico Santa Mar\'ia, Valpara\'iso; Chile.\\
% ...
$^{149}$Department of Physics, Technion, Israel Institute of Technology, Haifa; Israel.\\
$^{150}$Raymond and Beverly Sackler School of Physics and Astronomy, Tel Aviv University, Tel Aviv; Israel.\\
% ...
$^{a}$ Also Affiliated with an institute covered by a cooperation agreement with CERN.\\
$^{b}$ Also at An-Najah National University, Nablus; Palestine.\\
$^{c}$ Also at Borough of Manhattan Community College, City University of New York, New York NY; United States of America.\\
$^{d}$ Also at Bruno Kessler Foundation, Trento; Italy.\\
$^{e}$ Also at Center for High Energy Physics, Peking University; China.\\
$^{f}$ Also at Centro Studi e Ricerche Enrico Fermi; Italy.\\
% ...

\end{flushleft}


\end{document}